\documentclass[a4paper,11pt]{article}
\usepackage{cite}

\usepackage{amsmath}
\usepackage{amsfonts}
\usepackage{graphicx}
\usepackage{array}

\numberwithin{equation}{section}

\title{Using an Effective Charges Method to extract $\Lambda_{\overline{\text{MS}}}$ from event shape moments in $e^+e^-$ annihilation \vspace{1cm}}
\author{\vspace{1cm} C.J. Maxwell\footnote{email: \texttt{c.j.maxwell@durham.ac.uk}} , K.E. Morgan\footnote{email: \texttt{k.e.morgan@durham.ac.uk}}\\ \\  \textit{Institute for Particle Physics Phenomenology, University of Durham}, \\\textit{South Road, Durham, DH1 3LE, UK}\vspace{1cm}}
\date{}

\begin{document}

\maketitle

\large \vspace{-10cm}\vbox{\hbox{IPPP/11/49}
            \hbox{DCPT/11/98}}
\normalsize \vspace{9cm}

\begin{abstract}

\noindent
We use an Effective Charges (ECH) method to extract $\Lambda_{\overline{\text{MS}}}$, and hence $\alpha_{s}(M_{z})$, from event shape moments in $e^+e^-$ annihilation. We compare these results with ones obtained using standard $\overline{\text{MS}}$ perturbation theory. The ECH method at NLO is found to perform better than standard $\overline{\text{MS}}$ perturbation theory when applied to means of event shape observables. For example, when we apply the NLO ECH method to $\langle 1-T\rangle$ we get $\alpha_s(M_z)=0.1193\pm0.0003$. However ECH at NNLO is found to work less well than ECH at NLO, and the ECH method also fails to describe data for higher moments of event shapes. We attempt to explain this by considering the ECH $\beta$-function as an asymptotic series. We also examine the effect of adding two different models for non-perturbative power corrections to the perturbative approximation given by the ECH method and $\overline{\text{MS}}$ perturbation theory. Whilst only small power corrections are required when using ECH at NLO, it is found that these models are insufficient to counteract the undesirable behaviour of ECH at NNLO.

\end{abstract}

\section{Introduction}

In recent years much progress has been made in calculating coefficients for higher order terms in QCD perturbative expansions~\cite{Glover, GehrmannDeRidder, GehrmannDeRidder2}. This has enabled many QCD observables to be calculated to a greater accuracy~\cite{worldav, PDGreviewQCD, Luisoni2, Dissertori, Dissertori2}. These more accurate perturbative results can then be used to determine $\alpha_s(M_z)$ with increased precision~\cite{worldav, PDGreviewQCD, Luisoni2, Dissertori, Dissertori2}. However, the result of a calculation performed at any fixed order in perturbation theory depends on the renormalisation scheme (RS) and renormalisation scale, $\mu_R$, used. The standard way of dealing with this dependence is to choose a particular RS (a common choice is the modified minimal subtraction scheme, $\overline{\text{MS}}$~\cite{QCDreview}) and perform the calculation with $\mu_R$ set equal to some energy, Q. The scale Q is associated with the data in some way; for example, when using data from $e^+e^-$ collisions it is common to choose the centre of mass energy. A theoretical uncertainty is then assigned to the result by varying $\mu_R$ between $Q/2$ and $2Q$~\cite{Luisoni, PDGreviewQCD, QCDreview2}. The choices of $\mu_R$ and the upper and lower limits of its variation are completely arbitrary. This method can also result in theoretical uncertainties that are much greater than the experimental uncertainties involved. This treatment is therefore somewhat unsatisfactory.

An alternative approach is to use the Effective Charges (ECH) method \mbox{\!\!\cite{DharGupta, Grunberg, Burby, Grunberg2, Maxwell2, DinsdaleMaxwell, DharGupta2, Stevenson2, GuptaShirkovTarasov, Maxwell}}. This method integrates up a $\beta$-function type equation for the observable that is analogous to the running equation for the strong coupling constant. The resulting integration constant is related to $\Lambda_{\text{QCD}}$, the characteristic scale of QCD~\cite{PDGreviewQCD, QCDreview, QCDreview2}. $\Lambda_{\text{QCD}}$, known as the dimensional transmutation parameter, can be considered to be the truly fundamental parameter of QCD~\cite{DApaper}. As we shall see, the ECH method directly relates experimental observables with this parameter. Usually $\Lambda_{\overline{\text{MS}}}$ is considered. Once determined this underlying universal parameter of QCD can be used to evaluate $\alpha_s(M_z)$ using the QCD $\beta$-function equation. There is no unphysical dependence on $\mu_R$ and therefore no need to estimate a theoretical uncertainty to account for this dependence.

In this paper we analyse event shape moments, using both the ECH method and standard $\overline{\text{MS}}$ perturbation theory ($\overline{\text{MS}}$ PT), in order to extract $\alpha_s(M_z)$. The event shape moments have been calculated using data taken from $e^+e^-$ collisions at LEP (DELPHI, OPAL and L3) and PETRA (JADE)~\cite{OPAL, JADE, Pahl, DELPHIdata, Reinhardt, L3}. The DELPHI collaboration performed a similar study~\cite{DELPHI}, applying the ECH method and $\overline{\text{MS}}$ PT at next-to-leading order (NLO) to event shape means. We replicate this analysis, and also extend the work to next-to-NLO (NNLO) using the NNLO perturbative coefficients for event shape moments that have recently been calculated by Gehrmann-De Ridder \textit{et al.} in~\cite{Glover}. In a further extension of the DELPHI collaboration's analysis, we study some higher moments of event shape observables: specifically the first three moments of $1-T$, $C$, $B_W$, $B_T$ and $Y_3$ as well as the mean of $\rho_E$ (see Section \ref{Event Shape Obs} for details). As in~\cite{DELPHI} we also add power corrections to these perturbative models to determine whether the estimation of $\alpha_s$ is improved. We use the dispersive power correction model, outlined in references~\cite{Luisoni, DokshitzerWebber, SalamWicke}, and also a simple power correction model~\cite{DELPHI, CampbellGloverMaxwell}.

The plan of the paper is as follows. We start by outlining the ECH method in Section \ref{Theory}. We then introduce event shape observables and give more details of the data used in Section \ref{Event Shape Obs}. Our results are given in Section \ref{Results}. The values of $\alpha_s(M_z)$ extracted from the event shape moments using the ECH method and standard $\overline{\text{MS}}$ PT, at both NLO and NNLO, are given in Section \ref{Finding alpha}. In Section \ref{ECH plots} we show how the ECH method can be used to approximate data by working backwards from a fixed value of $\alpha_s(M_z)$. We provide a commentary of these results in Section \ref{Commentary1}. In Section \ref{rho(R)} we interpret some of the results by looking at the perturbative expansions of the event shape moments and values of the $\Lambda$ parameter (see also Section \ref{Theory}). Power corrections are then added to the ECH method and $\overline{\text{MS}}$ PT: fits for $\alpha_s(M_z)$ and a dispersive power correction parameter, $\alpha_0$, are performed and the results given in Section \ref{pcs}. Fit results for $\alpha_s(M_z)$ and a simple power correction parameter, $\kappa_0$, are given in Section \ref{simple pcs}. Another commentary is given at this point, in Section \ref{Commentary2}. Our conclusions are given in Section \ref{Conclusions}.

\section{Theory}\label{Theory}

Here we shall briefly derive the ECH method~\cite{DharGupta, Grunberg, Burby, Grunberg2, Maxwell2, DinsdaleMaxwell, DharGupta2, Stevenson2, GuptaShirkovTarasov, Maxwell} and show how it can be used to extract $\Lambda_{\overline{\text{MS}}}$ and hence the strong coupling constant. We shall start from the principle of dimensional analysis~\cite{DApaper}. Let us consider a dimensionless QCD observable which depends on a single energy scale, $\mathcal{R}(Q)$. Since the observable is dimensionless we have to introduce another, dimensionful, quantity $\Lambda$:
\begin{equation}\label{R=f(L/Q)}
 \mathcal{R}(Q)=f\left(\frac{\Lambda}{Q}\right)\;,
\end{equation}
where $f$ is some function of the dimensionful quantity $\Lambda$ and the physical scale $Q$. We will assume massless quarks; the extension to the case with non-zero quark mass has been considered in Refs \cite{GuptaShirkovTarasov,DinsdaleMaxwell}. $\Lambda$ must therefore be related to $\Lambda_\text{QCD}$, the fundamental dimensional transmutation scale of QCD. We want to extract $\Lambda$, and to do this we must invert \eqref{R=f(L/Q)}. To find the inverse function $f^{-1}$ we shall start by looking at the derivative of $\mathcal{R}(Q)$ with respect to $Q$:
\begin{equation}\label{dR/dQ}
 \frac{d\mathcal{R}(Q)}{dQ}\equiv\frac{\rho(\mathcal{R}(Q))}{Q}\;.
\end{equation}
We have again used dimensional analysis to determine the form of \eqref{dR/dQ}. It can be rewritten as:
\begin{equation}\label{dR/dlnQ}
 \frac{d\mathcal{R}(Q)}{d\ln Q}\equiv\rho(\mathcal{R}(Q))\;.
\end{equation}
To solve \eqref{dR/dlnQ} we need to know something about $\mathcal{R}(Q)$. We shall take the form of its perturbative expansion to be:
\begin{equation}\label{R=}
 \mathcal{R}(Q)={a}(1+r_1 {a}+r_2{a}^2+\dots)\;,
\end{equation}
where ${a}\equiv\alpha_s(\mu_R)/\pi$ and $\mu_R$ is the renormalisation scale. For any observable of this type, i.e. one that depends on only a single energy scale, we can always ensure that the perturbative expansion takes the form of \eqref{R=} by applying an appropriate scaling and, if necessary, raising to a power. This step therefore results in no loss of generality.

An observable with an expansion of the form of \eqref{R=} is known as an effective charge~\cite{Maxwell, Hamacher} as it has a running equation like the strong coupling does. The renormalisation group equation for $\alpha_s$, rewritten in terms of $a$, is: 
\begin{equation}\label{beta fn}
 \frac{\partial a}{\partial \ln{\mu_R}}\equiv\beta(a)=-b a^2(1+ca+\sum_{n=2}^\infty c_n a^n)\;.
\end{equation}
$\beta(a)$ denotes the QCD $\beta$-function. The first two coefficients, \mbox{$b\!=\!(33\!-\!2N_f)/6$} and \mbox{$c\!=\!(153\!-\!19N_f)/12b$}, are universal. The higher $c_n$ are scheme-dependent coefficients. In the $\overline{\text{MS}}$ scheme \mbox{$c_2^{\overline{\text{MS}}}=(77139-15099N_f+325N_f^2)/1728b$} \mbox{\!\!\cite{PDGreviewQCD, QCDreview}}. We can derive the analogous expression for the effective charge by setting $\mu_R=Q$ in \eqref{R=} and differentiating with respect to $\ln Q$. If we also invert \eqref{R=} to get an expression for $a(\mathcal{R})$ and substitute in we find that:
\begin{equation}\label{dR/dlnx}
 \frac{d\mathcal{R}(Q)}{d\ln Q}\equiv\rho(\mathcal{R}(Q))=-b\mathcal{R}^2(1+c\mathcal{R}+\sum_{n=2}^\infty \rho_n \mathcal{R}^n)\;.
\end{equation}
Again, $b$ and $c$ are the first two universal coefficients from \eqref{beta fn}. The $\rho_n$ are specific to the observable $\mathcal{R}(Q)$; they are RS invariant and $Q$-independent combinations of coefficients from \eqref{R=} and \eqref{beta fn}. For example:
\begin{equation}
 \rho_2=r_2+c_2-r_1 c-r_1^2\;.
\end{equation}

We now integrate up \eqref{dR/dlnx} using asymptotic freedom as a boundary condition, i.e. requiring that $\mathcal{R}(Q)\rightarrow0$ as $Q\rightarrow \infty$:
\begin{equation}\label{int1}
 \ln \frac{Q}{\Lambda_{\mathcal{R}}}=\int_0^{\mathcal{R}(Q)} \frac{dx}{\rho(x)}+\kappa\;.
\end{equation}
This step has introduced an integration constant that we have split into two parts: $\Lambda_\mathcal{R}$ and $\kappa$. $\Lambda_{\mathcal{R}}$ is a finite dimensionful scale specific to the observable $\mathcal{R}$. $\kappa$ is an infinite constant which must cancel the singularity in $\rho(x)$ at $x=0$. Once this requirement is met there is still some freedom to choose the exact form of $\kappa$. Any variation in the definition of $\kappa$ can be absorbed into the definition of $\Lambda_\mathcal{R}$. We choose $\kappa$ to be:
\begin{equation}
 \kappa=\int_0^{\infty} \frac{dx}{b x^2(1+cx)}\;.
\end{equation}

Adding in this explicit form for $\kappa$ we can now rewrite \eqref{int1} as:
\begin{equation}\label{int2}
 \begin{split}
   b \ln \frac{Q}{\Lambda_{\mathcal{R}}}
&=\int_{\mathcal{R}(Q)}^\infty \frac{dx}{x^2(1+c x)}+\int_0^{\mathcal{R}(Q)} dx \left[\frac{b}{\rho(x)}+\frac{1}{x^2(1+cx)}\right]\\
&=\text{F}(\mathcal{R})+\text{G}(\mathcal{R})\;.
 \end{split}
\end{equation}
The integration in $\kappa$ has been split into two parts: from $0$ to $\mathcal{R}(Q)$ and from $\mathcal{R}(Q)$ to $\infty$. Integrating up the first term on the right hand side we get:
 \begin{equation}
  \text{F}(\mathcal{R})=\frac{1}{\mathcal{R}}+c\ln\left[\frac{c\mathcal{R}}{1+c\mathcal{R}}\right]\;.
 \end{equation}
The form of the second integral, $\text{G}(\mathcal{R})$, depends on the order at which $\rho(x)$ is truncated. For example, at NLO we have \mbox{$\rho(x)=-b x^2(1+cx)$} which cancels with the \smash{$\frac{1}{x^2(1+cx)}$} term to give $\text{G}(\mathcal{R})=0$. At NNLO we have:
\begin{equation}
 \text{G}(\mathcal{R})=\int_0^{\mathcal{R}(Q)}dx\frac{\rho_2}{(1+cx)(1+cx+\rho_2x^2)}\:.
\end{equation}
This integration has an analytic result with three different forms, depending on the relative sizes of the coefficients $c$ and $\rho_2$:
\begin{align}
 &(i)\,4\rho_2>c^2;\qquad \Delta \equiv \sqrt{4\rho_2-c^2}\;, \nonumber \\
&\text{G}(\mathcal{R})=\frac{c}{2}\ln\left|\frac{(1+c\mathcal{R})^2}{1+c\mathcal {R}+\rho_2\mathcal{R}^2}\right|\!+\!\frac{2\rho_2-c^2}{\Delta}\!\left[\arctan \left[\frac{2\rho_2\mathcal{R}+c}{\Delta}\right]\!-\arctan \bigg[\frac{c}{\Delta}\bigg]\right], \nonumber \\[1cm]
&(ii)\,4\rho_2<c^2;\qquad \Delta \equiv \sqrt{c^2-4\rho_2}\;, \nonumber \\
&\text{G}(\mathcal{R})=\frac{c}{2}\ln \left| \frac{(1+c\mathcal{R})^2}{1+c\mathcal{R}+\rho_2\mathcal{R}^2}\right|+\frac{2\rho_2-c^2}{2\Delta}\left[\ln\left|\frac{2\rho_2\mathcal{R}+c-\Delta}{2\rho_2\mathcal{R}+c+\Delta}\right|+\ln\left|\frac{c+\Delta}{c-\Delta} \right|\right], \nonumber \\[1cm]
&(iii)\,4\rho_2=c^2; \nonumber \\
&\text{G}(\mathcal{R})=\frac{c}{2}\ln\left|\frac{(1+c\mathcal{R})^2}{1+c\mathcal{R}+\rho_2\mathcal{R}^2}\right|+(c^2-2\rho_2)\left[\frac{1}{2\rho_2\mathcal{R}+c}-\frac{1}{c}\right]\;.
\end{align}

We now rearrange for $\Lambda_\mathcal{R}$ in \eqref{int2}:
\begin{equation}\label{Lambda=}
 \Lambda_\mathcal{R}=Q\mathcal{F}(\mathcal{R})\mathcal{G}(\mathcal{R})\;,
\end{equation}
where
\begin{equation}\label{curly F}
\mathcal{F}(\mathcal{R})=\exp \left(-\frac{\text{F}(\mathcal{R})}{b}\right)=e^{-1/b\mathcal{R}}{\left(1+\frac{1}{c\mathcal{R}} \right)}^{c/b}\;, 
\end{equation}
and $\mathcal{G}(\mathcal{R})= \exp \left(-\text{G}(\mathcal{R})/b\right)$. $\Lambda_\mathcal{R}$ is a scheme independent quantity associated with the specific effective charge in question. It can be related to the universal quantity $\tilde{\Lambda}_{\overline{\text{MS}}}$ using:
\begin{equation}\label{Lambda twiddle}
 \tilde{\Lambda}_{\overline{\text{MS}}}=\Lambda_{\mathcal{R}}\, e^{-r/b}\;,
\end{equation}
where $r\equiv r_1^{\overline{\text{MS}}}(\mu_R=Q)$. The tilde here indicates that this definition is slightly different from the standard one for $\Lambda_{\overline{\text{MS}}}$ ($\Lambda_{\text{QCD}}$ in the $\overline{\text{MS}}$ scheme). This is due to our choice of $\kappa$ and, to regain the standard definition of
$\Lambda_{\overline{\text{MS}}}$~\cite{kappashift, kappashift2}, we have to shift $\kappa$ by $c\ln(b/2c)$ resulting in the relation:
\begin{align}\label{Lambda MSbar}
 \Lambda_{\overline{\text{MS}}}&=\left(\frac{2c}{b}\right)^{c/b}\tilde{\Lambda}_{\overline{\text{MS}}} \nonumber \\
&=\left(\frac{2c}{b}\right)^{c/b} Q\mathcal{F}(\mathcal{R})\mathcal{G}(\mathcal{R}) e^{-r/b}\;.
\end{align}
Note that all scheme dependence in $\Lambda_{\overline{\text{MS}}}$ is contained in the exponential factor $e^{-r/b}$. Although $\Lambda_{\overline{\text{MS}}}$ is a scheme dependent quantity it is possible to relate $\Lambda$'s in different schemes exactly, providing that the NLO coefficients are known in both schemes. One simply notes that, from \eqref{Lambda twiddle}, \mbox{$\tilde{\Lambda}_{\overline{\text{MS}}}\, e^{r/b} =\Lambda_{\mathcal{R}}$}. Since $\Lambda_{\mathcal{R}}$ is scheme invariant we can equate the left-hand sides of the expression in two different schemes. For instance for $\Lambda$ in the $\overline{\text{MS}}$ scheme and $\Lambda$ in the MS scheme the relation is:
\begin{equation*}
\Lambda_{\overline{\text{MS}}}=\text{exp}\left\{\frac{r^{\text{MS}}-r^{\overline{\text{MS}}}}{b}\right\} \Lambda_{\text{MS}}\;,
\end{equation*}
where $r^{\overline{\text{MS}}}$ and $r^{\text{MS}}$ are the NLO coefficients in the $\overline{\text{MS}}$ and MS schemes respectively.

From $\Lambda_{\overline{\text{MS}}}$ we can find $\alpha_s(M_z)$ using:
\begin{align}\label{alpha=}
 a&=\frac{-1}{c\left[1+W\left(-\exp\left[-\left(\displaystyle1+\frac{b}{c}\ln\frac{M_z}{\Huge{\tilde{\Lambda}_{\overline{\text{MS}}}}}\right)\right]\right)\right]}\;,\\
\alpha_s&=\begin{cases} \pi a & \text{for NLO} \label{alpha2=}\\
           \pi(a+c_2 a^3) & \text{for NNLO}\;,
          \end{cases}
\end{align}
where $W$ is the Lambert $W$-function~\cite{PApaper} defined by \mbox{$W(z) \exp (W(z))=z$}. The $W_{-1}$ branch should be used to ensure asymptotic freedom~\cite{Corless}. The coupling $a$ in \eqref{alpha=} is the ${\overline{\text{MS}}}$ coupling corresponding to a two-loop $\beta$-function solution, with the $\beta$-function coefficients $c_2$ and higher set to zero. This can be related to the coupling needed at NNLO, which has a non-zero $c_2$ but vanishing higher coefficients, by an expansion in powers of $a$. The advantage of this technique is that $a$ at two-loops can be written analytically, in terms of the Lambert $W$-function, and one avoids having to solve transcendental equations.

\section{Event Shape Observables}\label{Event Shape Obs}

Event shape observables give information about the topology of an event, and are designed to be IR- and UV-safe. They are good observables to use to study $\alpha_s$ since they are sensitive to the addition of extra final state partons, and hence to extra orders of $\alpha_s$~\cite{PDGreviewQCD}. In this paper we analyse event shape moments. The $n^\text{th}$ moment of an event shape observable, $y$, is given by~\cite{Luisoni, Glover, DELPHI}:
\begin{equation}\label{yn}
 \langle y^n\rangle=\frac{1}{\sigma_\text{had}} \int_0^{y_{\text{max}}}y^n\frac{d\sigma}{dy}dy\;,
\end{equation}
where $y_{\text{max}}$ is the greatest value of the event shape that is kinematically allowed and $\sigma_{\text{had}}$ is the hadronic cross-section.

$\langle y^n\rangle$ has the following perturbative expansion~\cite{Luisoni, Glover}:
\begin{align}\label{pt exp y}
 \langle y^n\rangle=&
\left(\frac{\alpha_s(\mu_R)}{2\pi}\right)\overline{\mathcal{A}}_{y,n} +\left(\frac{\alpha_s(\mu_R)}{2\pi}\right)^2\left(\overline{\mathcal{B}}_{y,n} +\overline{\mathcal{A}}_{y,n}\,b\ln\frac{\mu_R^2}{Q^2}\right) \nonumber\\
&+\left(\frac{\alpha_s(\mu_R)}{2\pi}\right)^3\!\left(\overline{\mathcal{C}}_{y,n} +2\overline{\mathcal{B}}_{y,n}\,b\ln\frac{\mu_R^2}{Q^2} +\overline{\mathcal{A}}_{y,n}\! \left(b^2\ln^2\frac{\mu_R^2}{Q^2}+2bc\ln\frac{\mu_R^2}{Q^2}\right)\right) \nonumber\\
&+\ldots\;,
\end{align}
where $b$ and $c$ are coefficients from the running coupling equation \eqref{beta fn}, and $\overline{\mathcal{A}}_{y,n}$, $\overline{\mathcal{B}}_{y,n}$ and $\overline{\mathcal{C}}_{y,n}$ are perturbative coefficients that are specific to the event shape moment. The NNLO perturbative coefficients ($\overline{\mathcal{C}}_{y,n}$) for several event shapes have recently been calculated~\cite{Glover}.

In order to apply the ECH method to an event shape moment, we must first rearrange \eqref{pt exp y} to get it into the form of an effective charge as given in \eqref{R=}. To do this we divide through by $\overline{\mathcal{A}}_{y,n}$ and, remembering that $a\equiv\alpha_s(\mu_R)/\pi$, put in appropriate factors of two:
\begin{equation}\label{R to y}
 \mathcal{R}(Q)=\frac{2\langle y^n\rangle}{\overline{\mathcal{A}}_{y,n}}= a\left(1+\frac{\overline{\mathcal{B}}_{y,n}}{2\overline{\mathcal{A}}_{y,n}} a+\frac{\overline{\mathcal{C}}_{y,n}}{4\overline{\mathcal{A}}_{y,n}} a^2+\dots\right)\;.
\end{equation}
For clarity we have taken $\mu_R=Q$. Comparing this with \eqref{R=} we see that
\begin{equation}
r_1=\frac{\overline{\mathcal{B}}_{y,n}}{2\overline{\mathcal{A}}_{y,n}} \qquad
\text{and} \qquad
r_2=\frac{\overline{\mathcal{C}}_{y,n}}{4\overline{\mathcal{A}}_{y,n}}\;.
\end{equation}

In this paper we examine the first three moments of five different event shape observables: one minus the thrust ($1-T$)~\cite{thrust, thrust2}, the C-parameter ($C$)~\cite{c, c2}, the wide and total jet broadenings ($B_W$ and $B_T$)~\cite{broadening, broadening2}, and the three-to-two jet transition parameter in the Durham algorithm ($Y_3$)~\cite{y3}. In addition, we analyse the mean of the heavy jet mass in the E-scheme ($\rho_E=(M_h^2/s)_E$)~\cite{rho, DELPHI}.

The data come from various LEP experiments including OPAL~\cite{OPAL, Pahl}, DELPHI~\cite{DELPHIdata, Reinhardt} and L3~\cite{L3}. Some low energy JADE (PETRA) data are also included~\cite{JADE, Pahl}. The data cover a range of energies from \mbox{14.0\,-\,206.6 GeV}.

\begin{figure}
 \begin{center}
  \input{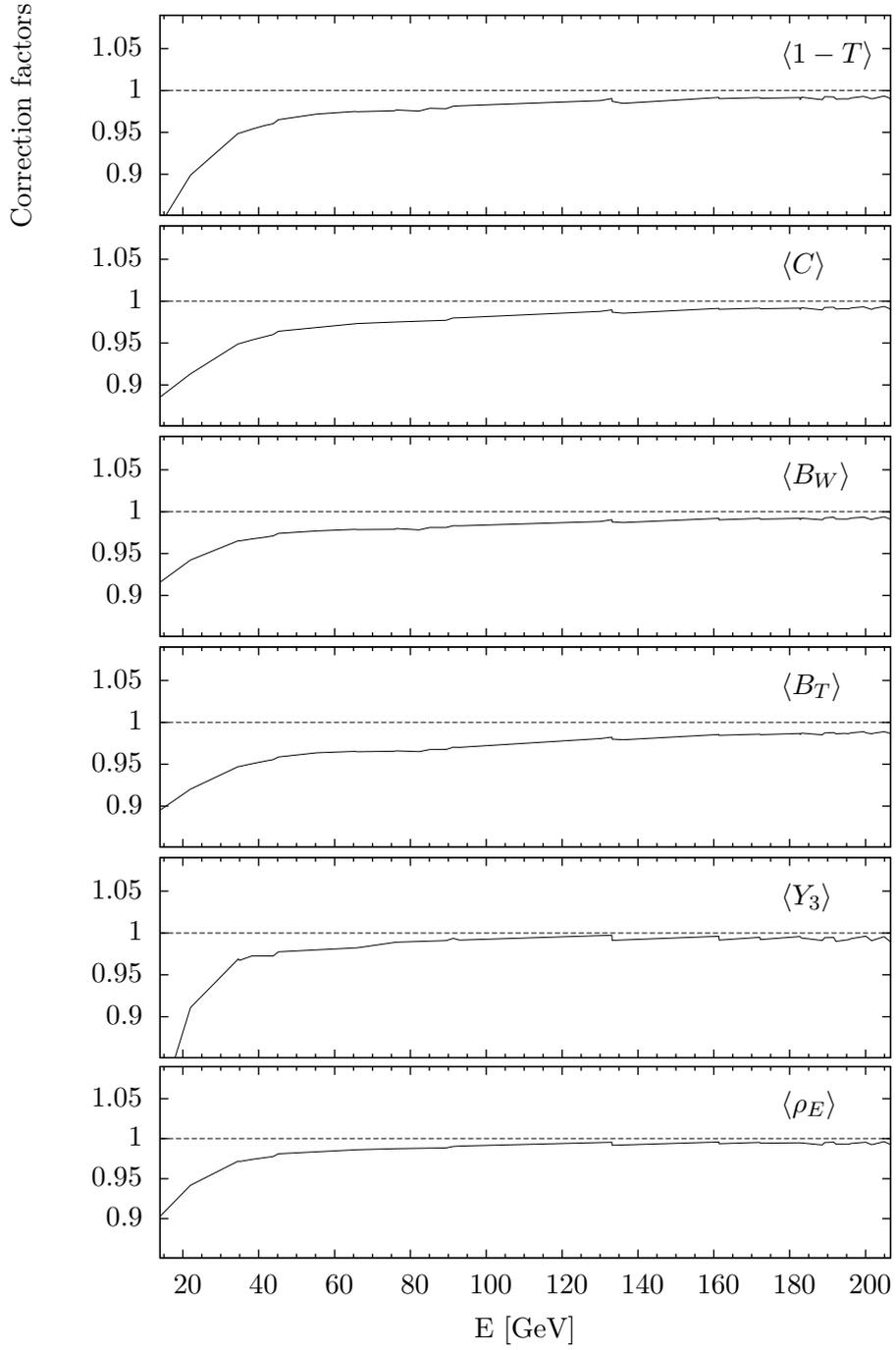}
\caption{Heavy quark mass correction factors for the event shape means}
\label{corrfactors}
 \end{center}
\end{figure}

All observables are corrected for b and c decays using HERWIG++ simulations~\cite{Herwig,Herwig2}. Samples of $10^6$ events were run with $N_f=3$ (only light quarks) and also with $N_f=5$ (with b and c quarks). The event shape moment was calculated in both of these cases and a ratio of the two quantities was taken to give a correction factor. This was done for each energy. The data in this paper have been multiplied by the appropriate correction factors. The correction factors for the means ($n=1$ moments) are shown in Fig. \ref{corrfactors}. Similar trends are seen in the $n=2$ and 3 moments, although the deviation of the correction factors from 1 tends to get larger at low energies for higher moments. These corrections have been compared with those calculated by the DELPHI collaboration in Fig. 3 of~\cite{DELPHI}. The factors in Fig. \ref{corrfactors} show the same energy dependence as in~\cite{DELPHI}, although our correction factors deviate further from 1 at low energies. This is because we have corrected for c decays as well as b decays. When only b decays are taken into consideration the corrections we calculated using HERWIG++ agree well with those that the DELPHI collaboration calculated using PYTHIA 6.1.

\section{Results}\label{Results}

\subsection{Finding $\tilde{\Lambda}_{\overline{\text{MS}}}$ and $\alpha_s(M_z)$ from data}\label{Finding alpha}

Now that we have corrected data in the form of an effective charge (see Section \ref{Event Shape Obs}) we can use the ECH method to extract $\alpha_s(M_z)$. For every event shape moment, and at each energy, a value of $\tilde{\Lambda}_{\overline{\text{MS}}}$ is obtained using \eqref{Lambda=} and \eqref{Lambda twiddle}. An overall value of $\tilde{\Lambda}_{\overline{\text{MS}}}$ is then found by performing a weighted average over all the energies for a specific event shape moment. The corresponding values of $\alpha_s(M_z)$ are calculated from the averages using \eqref{alpha=} and \eqref{alpha2=}. This is done at both NLO and NNLO.

Table \ref{table1} shows the weighted averages of $\tilde{\Lambda}_{\overline{\text{MS}}}$ obtained from the means and the corresponding values of $\alpha_s(M_z)$. These results are also plotted in the top two panels of Fig. \ref{scatter alpha}.

For comparison, $\tilde{\Lambda}_{\overline{\text{MS}}}$ and $\alpha_s(M_z)$ are also extracted from the event shape means using $\overline{\text{MS}}$ PT and equation \eqref{pt exp y}. The results are shown in Table \ref{table MSbar} and the bottom half of Fig. \ref{scatter alpha}. The first uncertainty on the $\overline{\text{MS}}$ PT results is the experimental uncertainty combined in quadrature with the uncertainties from the perturbative coefficients $\overline{\mathcal{B}}_{y,n}$ and $\overline{\mathcal{C}}_{y,n}$~\cite{Glover}. 
These small uncertainties arise from the numerical integration of the perturbative coefficients associated with the differential cross section $\frac{d\sigma}{dy}$ in equation \eqref{yn}.
The second is a theoretical scale uncertainty obtained by varying $\mu_R$ between $\mu_R=0.5Q$ and $\mu_R=2Q$. The largest difference between the calculations using the central value of $\mu_R$ and the upper and lower variations is taken as the uncertainty. The error bars in Fig. \ref{scatter alpha} represent the combined experimental and perturbative coefficient uncertainty and the theoretical scale uncertainty added in quadrature. In contrast, the uncertainty on the ECH results is just the combined experimental and perturbative coefficient uncertainty. Results for the n=2 moments are shown in Tables \ref{n=2 alphas}-\ref{table MSbar n2} and Fig. \ref{scatter alpha n2}, and for the n=3 moments in Tables \ref{n=3 alphas}-\ref{table MSbar n3} and Fig. \ref{scatter alpha n3}.

{\renewcommand{\arraystretch}{1.5}
\renewcommand{\tabcolsep}{0.2cm}
\begin{table}
\small
\vspace{-1cm}
\begin{center}
\begin{tabular}{ccccccc}
\hline
\hline
Observable & & $\tilde{\Lambda}_{\overline{\text{MS}}}$ [MeV] & $\alpha_s(M_z)$ & & $\tilde{\Lambda}_{\overline{\text{MS}}}$ [MeV] & $\alpha_s(M_z)$\\
\hline
 & & \multicolumn{2}{c}{NLO} & & \multicolumn{2}{c}{NNLO}\\
\hline
$\langle 1-T \rangle$ & & $262\pm4$ & $0.1193\pm0.0003$ & & $319\pm6$ & $0.1233\pm0.0004$\\
$\langle C \rangle$ & & $269\pm4$ & $0.1198\pm0.0003$ & & $300\pm5$ & $0.1221\pm0.0003$\\
$\langle B_W\rangle$ & & $326\pm3$ & $0.1234\pm0.0002$ & & $312\pm4$ & $0.1229\pm0.0003$\\
$\langle B_T\rangle$ & & $265\pm2$ & $0.1196\pm0.0002$ & & $337\pm5$ & $0.1244\pm0.0003$\\
$\langle Y_3\rangle$ & & $176\pm5$ & $0.1126\pm0.0004$ & & $211\pm7$ & $0.1157\pm0.0006$\\
$\langle \rho_E\rangle$ & & $268\pm5$ & $0.1198\pm0.0003$ & & $289\pm7$ & $0.1214\pm0.0004$\\
\hline
\hline
\end{tabular}
\caption{Weighted averages for $\tilde{\Lambda}_{\overline{\text{MS}}}$ and the corresponding values of $\alpha_s(M_z)$ for the means using the ECH method. \label{table1}}
\vspace{3cm}
 \begin{tabular}{ccc@{$\, \pm \,$}c@{$\, \pm \,$}ccc@{$\, \pm \,$}c@{$\, \pm \,$}ccc}
  \hline
\hline
Observable & \hspace{0.1cm} &
\multicolumn{3}{c}{$\tilde{\Lambda}_{\overline{\text{MS}}}$ [MeV]} & \hspace{0.1cm} & \multicolumn{3}{c}{$\alpha_s(M_z)$} & \hspace{0.1cm} & $\chi^2$/dof\\
\hline
\multicolumn{11}{c}{NLO}\\
\hline
$\langle 1-T \rangle$ & & 646 & 9 & 279 & & 0.1385 & 0.0003 & 0.0107 & & 52.3/47\\
$\langle C \rangle$ & & 597 & 7 & 247 & & 0.1365 & 0.0003 & 0.0099 & & 45.7/41\\
$\langle B_W \rangle$ & & 336 & 3 & 105 & & 0.1241 & 0.0002 & 0.0062 & & 39.5/47\\
$\langle B_T \rangle$ & & 429 & 3 & 147 & & 0.1291 & 0.0001 & 0.0073 & & 69.7/47\\
$\langle Y_3 \rangle$ & & 268 & 5 & 81 & & 0.1197 & 0.0003 & 0.0056 & & 61.5/32\\
$\langle \rho_E \rangle$ & & 340 & 6 & 90 & & 0.1243 & 0.0004 & 0.0053 & & 11.5/14\\
\hline
\hline
\multicolumn{11}{c}{NNLO}\\
\hline
$\langle 1-T \rangle$ & & 396 & 7 & 77 & & 0.1277 & 0.0004 & 0.0041 & & 51.1/47\\
$\langle C \rangle$ & & 378 & 6 & 69 & & 0.1267 & 0.0003 & 0.0038 & & 44.3/41\\
$\langle B_W \rangle$ & & 317 & 12 & 13 & & 0.1232 & 0.0007 & 0.0008 & & 37.8/47\\
$\langle B_T \rangle$ & & 347 & 9 & 36 & & 0.1250 & 0.0005 & 0.0021 & & 78.1/47\\
$\langle Y_3 \rangle$ & & 228 & 5 & 18 & & 0.1171 & 0.0004 & 0.0014 & & 65.1/32\\
$\langle \rho_E \rangle$ & & 295 & 7 & 21 & & 0.1218 & 0.0005 & 0.0014 & & 11.5/14\\
\hline
\hline
 \end{tabular}
\caption{$\tilde{\Lambda}_{\overline{\text{MS}}}$ and $\alpha_s(M_z)$ for the means using $\overline{\text{MS}}$ PT. The first error is experimental and the second is theoretical.
\label{table MSbar}}
\end{center}
\end{table}

\begin{figure}
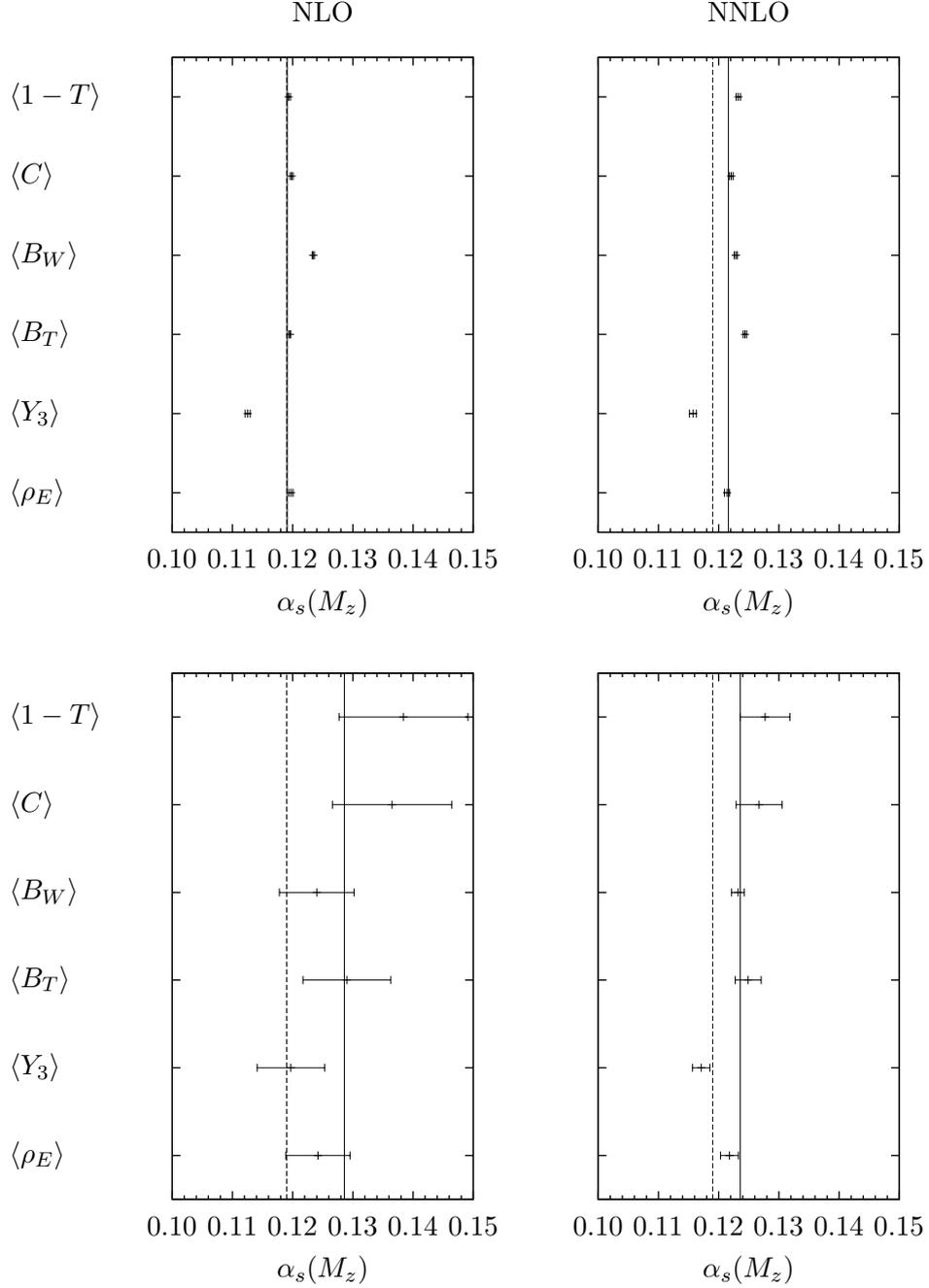

\begin{center}
\begin{tabular}{cc}
  \input{scatteralphaNLO.tex}
& \hspace{-1.75cm}
  \input{scatteralphaNNLO.tex}\\
  \input{scatteralphaNLOMSbar.tex}
& \hspace{-1.75cm}
  \input{scatteralphaNNLOMSbar.tex}
\end{tabular}
\vspace{-0.5cm}
\caption{Scatter plots showing the values of $\alpha_s(M_z)$ obtained from the means of the event shapes using the ECH method (top two panels) and $\overline{\text{MS}}$ PT (bottom two panels). The dotted lines show the value of the coupling obtained from N$^3$LO calculations on Z decays: $\alpha_s(M_z)=0.1190$\protect{~\cite{Baikov}}. The solid lines show the unweighted average for the NLO and NNLO cases.}
\label{scatter alpha}
\end{center}
\end{figure}

\begin{table}
\small
\vspace{-1cm}
\begin{center}
\begin{tabular}{ccc@{$\, \pm \,$}cc@{$\, \pm \,$}ccc@{$\, \pm \,$}cc@{$\, \pm \,$}c}
 \hline
\hline
Observable &  & \multicolumn{2}{c}{$\tilde{\Lambda}_{\overline{\text{MS}}}$ [MeV]} & \multicolumn{2}{c}{$\alpha_s(M_z)$} & & \multicolumn{2}{c}{\hspace{-0.2cm}$\tilde{\Lambda}_{\overline{\text{MS}}}$ [MeV]} & \multicolumn{2}{c}{$\alpha_s(M_z)$}\\
\hline
 & & \multicolumn{4}{c}{NLO} & & \multicolumn{4}{c}{NNLO}\\
\hline
$\langle (1-T)^2 \rangle$ & & 130 & 4 & 0.1079 & 0.0005 & & 438 & \hspace{-0.5cm} 22 & 0.1298 & 0.0011 \\
$\langle C^2 \rangle$ & & 123 & 3 & 0.1071 & 0.0004 & & 466 & \hspace{-0.5cm} 30 & 0.1312 & 0.0014 \\
$\langle B_W^2\rangle$ & & 166 & 5 & 0.1116 & 0.0005 & & 203 & \hspace{-0.5cm} 7 & 0.1151 & 0.0006 \\
$\langle B_T^2\rangle$ & & 0.05 & 0.02 & 0.0520 & 0.0017 & & 287 & \hspace{-0.5cm} 34 & 0.1213 & 0.0023 \\
$\langle Y_3^2\rangle$ & & 32 & 13 & 0.0905 & 0.0041 & & 38 & \hspace{-0.5cm} 16 & 0.0924 & 0.0045 \\
\hline
\hline
\end{tabular}
\caption{$\tilde{\Lambda}_{\overline{\text{MS}}}$ and $\alpha_s(M_z)$ for the n=2 moments using the ECH method. \label{n=2 alphas}}
\vspace{3cm}
 \begin{tabular}{cc@{$\, \pm \,$}c@{$\, \pm \,$}ccc@{$\, \pm \,$}c@{$\, \pm \,$}ccc}
  \hline
\hline
Observable & \multicolumn{3}{c}{$\tilde{\Lambda}_{\overline{\text{MS}}}$ [MeV]} & \hspace{0.1cm} & \multicolumn{3}{c}{$\alpha_s(M_z)$} & \hspace{0.1cm} & $\chi^2$/dof\\
\hline
\multicolumn{10}{c}{NLO}\\
\hline
$\langle (1-T)^2 \rangle$ & 1047 & 30 & 607 & & 0.1515 & 0.0008 & 0.0172 & & 23.6/32\\
$\langle C^2 \rangle$ & 1048 & 29 & 609 & & 0.1516 & 0.0008 & 0.0173 & & 28.3/26\\
$\langle B_W^2 \rangle$ & 245 & 9 & 76 & & 0.1181 & 0.0007 & 0.0055 & & 22.5/32\\
$\langle B_T^2 \rangle$ & 775 & 15 & 490 & & 0.1431 & 0.0005 & 0.0167 & & 26.9/32\\
$\langle Y_3^2 \rangle$ & 184 & 29 & 63 & & 0.1132 & 0.0026 & 0.0056 & & 45.5/17\\
\hline
\hline
\multicolumn{10}{c}{NNLO}\\
\hline
$\langle (1-T)^2 \rangle$ & 467 & 13 & 147 & & 0.1312 & 0.0006 & 0.0070 & & 23.1/32\\
$\langle C^2 \rangle$ & 469 & 13 & 147 & & 0.1313 & 0.0006 & 0.0070 & & 26.4/26\\
$\langle B_W^2 \rangle$ & 207 & 8 & 19 & & 0.1154 & 0.0006 & 0.0015 & & 22.5/32\\
$\langle B_T^2 \rangle$ & 370 & 7 & 110 & & 0.1263 & 0.0004 & 0.0061 & & 26.2/32\\
$\langle Y_3^2 \rangle$ & 147 & 23 & 16 & & 0.1099 & 0.0024 & 0.0017 & & 46.9/17\\
\hline
\hline
 \end{tabular}
\caption{$\tilde{\Lambda}_{\overline{\text{MS}}}$ and $\alpha_s(M_z)$ for the n=2 moments using $\overline{\text{MS}}$ PT. The first error is experimental and the second is theoretical.
\label{table MSbar n2}}
\end{center}
\end{table}

\begin{figure}
 \begin{center}
  \begin{tabular}{cc}
\vspace{-0.75cm}
   \input{scatteralphaNLOn2.tex} & \hspace{-1.75cm}
\input{scatteralphaNNLOn2.tex}\\
\input{scatteralphaNLOMSbarn2.tex}
& \hspace{-1.75cm}
  \input{scatteralphaNNLOMSbarn2.tex}
  \end{tabular}
\vspace{-0.5cm}
\caption{Scatter plots showing the values of $\alpha_s(M_z)$ obtained from the n=2 moments of the event shapes using the ECH method (top two panels) and $\overline{\text{MS}}$ PT (bottom two panels). The solid and dotted lines are as in Fig \ref{scatter alpha}.}
\label{scatter alpha n2}
 \end{center}
\end{figure}

\begin{table}
\small
\vspace{-1cm}
\begin{center}
\begin{tabular}{ccc@{$\, \pm \,$}cc@{$\, \pm \,$}ccc@{$\, \pm \,$}cc@{$\, \pm \,$}c}
 \hline
\hline
Observable &  & \multicolumn{2}{c}{$\tilde{\Lambda}_{\overline{\text{MS}}}$ [MeV]} & \multicolumn{2}{c}{$\alpha_s(M_z)$} & & \multicolumn{2}{c}{\hspace{-0.2cm}$\tilde{\Lambda}_{\overline{\text{MS}}}$ [MeV]} & \multicolumn{2}{c}{$\alpha_s(M_z)$}\\
\hline
 & & \multicolumn{4}{c}{NLO} & & \multicolumn{4}{c}{NNLO}\\
\hline
$\langle (1-T)^3 \rangle$ & & 90 & \hspace{-0.5cm} 5 & 0.1027 & 0.0007 & & 647 & \hspace{-0.5cm} 37 & 0.1389 & 0.0014 \\
$\langle C^3 \rangle$ & & 73 & \hspace{-0.5cm} 3 & 0.0999 & 0.0006 & & 344 & \hspace{-0.5cm} 57 & 0.1248 & 0.0033 \\
$\langle B_W^3\rangle$ & & 117 & \hspace{-0.5cm} 6 & 0.1063 & 0.0007 & & 152 & \hspace{-0.5cm} 7 & 0.1104 & 0.0007 \\
$\langle B_T^3\rangle$ & & 7.5 & \hspace{-0.5cm} 0.5 & 0.0776 & 0.0005 & & 49 & \hspace{-0.5cm} 8 & 0.0953 & 0.0018 \\
$\langle Y_3^3 \rangle$ & & 0.6 & \hspace{-0.5cm} 1.4 & 0.0620 & 0.0122 & & 0.6 & \hspace{-0.5cm} 1.6 & 0.0624 & 0.0131 \\
\hline
\hline
\end{tabular}
\caption{$\tilde{\Lambda}_{\overline{\text{MS}}}$ and $\alpha_s(M_z)$ for the n=3 moments using the ECH method \label{n=3 alphas}}
\vspace{3cm}
 \begin{tabular}{cc@{$\, \pm \, $}c@{$\, \pm \, $}ccc@{$\, \pm \,$}c@{$\, \pm \,$}ccc}
  \hline
\hline
Observable & \multicolumn{3}{c}{$\tilde{\Lambda}_{\overline{\text{MS}}}$ [MeV]} & \hspace{0.1cm} & \multicolumn{3}{c}{$\alpha_s(M_z)$} & \hspace{0.1cm} & $\chi^2$/dof\\
\hline
\multicolumn{10}{c}{NLO}\\
\hline
$\langle (1-T)^3 \rangle$ & 1249 & 35 & 780 & & 0.1570 & 0.0009 & 0.0200 & & 7.4/17\\
$\langle C^3 \rangle$ & 1316 & 33 & 844 & & 0.1587 & 0.0008 & 0.0210 & & 9.4/17\\
$\langle B_W^3 \rangle$ & 177 & 10 & 56 & & 0.1126 & 0.0009 & 0.0051 & & 8.2/17\\
$\langle B_T^3 \rangle$ & 976 & 21 & 692 & & 0.1495 & 0.0006 & 0.0205 & & 9.1/17\\
$\langle Y_3^3 \rangle$ & 163 & 40 & 55 & & 0.1113 & 0.0039 & 0.0053 & & 67.2/17\\
\hline
\hline
\multicolumn{10}{c}{NNLO}\\
\hline
$\langle (1-T)^3 \rangle$ & 493 & 14 & 176 & & 0.1324 & 0.0006 & 0.0081 & & 7.2/17\\
$\langle C^3 \rangle$ & 526 & 14 & 191 & & 0.1340 & 0.0006 & 0.0084 & & 10.0/17\\
$\langle B_W^3 \rangle$ & 151 & 8 & 13 & & 0.1103 & 0.0008 & 0.0013 & & 8.5/17\\
$\langle B_T^3 \rangle$ & 400 & 9 & 141 & & 0.1279 & 0.0005 & 0.0074 & & 9.7/17\\
$\langle Y_3^3 \rangle$ & 135 & 33 & 13 & & 0.1086 & 0.0037 & 0.0015 & & 68.5/17\\
\hline
\hline
 \end{tabular}
\caption{$\tilde{\Lambda}_{\overline{\text{MS}}}$ and $\alpha_s(M_z)$ for the n=3 moments using $\overline{\text{MS}}$ PT. The first error is experimental and the second is theoretical. \label{table MSbar n3}}
\end{center}
\end{table}

\begin{figure}
 \begin{center}
  \begin{tabular}{cc}
\vspace{-0.75cm}
   \input{scatteralphaNLOn3.tex} & \hspace{-1.75cm}
\input{scatteralphaNNLOn3.tex}\\
\input{scatteralphaNLOMSbarn3.tex}
& \hspace{-1.75cm}
  \input{scatteralphaNNLOMSbarn3.tex}
  \end{tabular}
\vspace{-0.5cm}
\caption{Scatter plots showing the values of $\alpha_s(M_z)$ obtained from the n=3 moments of the event shapes using the ECH method (top two panels) and $\overline{\text{MS}}$ PT (bottom two panels). The solid and dotted lines are as in Fig \ref{scatter alpha}.}
\label{scatter alpha n3}
 \end{center}
\end{figure}

\subsection{Approximating data from a fixed value of $\alpha_s(M_z)$}\label{ECH plots}

Another way to implement the ECH method is to work backwards from a fixed value of $\alpha_s(M_z)$ to get an approximation to the data. In this section we fix $\alpha_s(M_z)$ at the result extracted from next-to-NNLO (N$^3$LO) calculations on Z-decays~\cite{Baikov}: $\alpha_s(M_z)=0.1190\pm 0.0026$. This value is close to the world average value of $\alpha_s(M_z)=0.1184\pm0.0031$ reported in \cite{worldav}. The N$^3$LO Z-decay value corresponds to \mbox{$\tilde{\Lambda}_{\overline{\text{MS}}}=254$ MeV}. We can then find $\Lambda_\mathcal{R}$ using \eqref{Lambda twiddle} and hence $\mathcal{R}(Q)$ by inverting \eqref{Lambda=}. At NLO, where $\mathcal{G} (\mathcal{R})=1$, this can be done analytically:
\begin{equation} \label{R(Q)}
 \mathcal{R}_{\text{NLO}}(Q)=\frac{-1}{c\left[1+W\left(-\exp\left[-\left(\displaystyle1+\frac{b}{c} \ln\frac{Q}{\Lambda_{\mathcal{R}}}\right)\right]\right)\right]}\;.
\end{equation}
Note that if we rewrite ${\Lambda_{\mathcal{R}}}$ in terms of $\tilde{\Lambda}_{\overline{\text{MS}}}$ using \eqref{Lambda twiddle} and compare it with \eqref{alpha=} we find that $\mathcal{R}_{\text{NLO}}(Q)=\alpha_s\,(\mu_{\text{ECH}})/\pi$. The particular choice of $\overline{\text{MS}}$ renormalisation scale of $\mu_{\text{R}}=\mu_{\text{ECH}}=e^{-r/b}\,Q$ is equivalent to the ECH scheme. This ECH scale corresponds to absorbing radiative corrections into the definition of the coupling. 

A NLO ECH approximation to data is then given by \mbox{$\langle y^n\rangle_{\text{NLO}}\!=\!\frac{\overline{\mathcal{A}}_{y,n}}{2}\mathcal{R}_{\text{NLO}}$}, using \eqref{R to y}. At NNLO it is not possible to invert \eqref{Lambda=} analytically and so this is done numerically using Mathematica 7.0~\cite{Mathematica}. 

\begin{figure}
 \begin{center}
  \begin{tabular}{c}
\vspace{-0.5cm}
   \input{echTn1.tex}\\
\vspace{-0.5cm}
   \input{echTn2.tex}\\
\vspace{-0.5cm}
   \input{echTn3.tex}\\
  \end{tabular}
\caption{ECH and $\overline{\text{MS}}$ PT approximations to data at NLO and NNLO for \mbox{$\langle (1-T)^n\rangle$}, with n=1,2,3 from top to bottom.}
\label{ECH plot T}
 \end{center}
\end{figure}

\begin{figure}
 \begin{center}
  \begin{tabular}{c}
\vspace{-0.5cm}
   \input{echCn1.tex}\\
\vspace{-0.5cm}
   \input{echCn2.tex}\\
\vspace{-0.5cm}
   \input{echCn3.tex}\\
  \end{tabular}
\caption{ECH and $\overline{\text{MS}}$ PT approximations to data at NLO and NNLO for $\langle C^n\rangle$, with n=1,2,3 from top to bottom.}
\label{ECH plot C}
 \end{center}
\end{figure}

\begin{figure}
 \begin{center}
  \begin{tabular}{c}
\vspace{-0.5cm}
   \input{echBWn1.tex}\\
\vspace{-0.5cm}
   \input{echBWn2.tex}\\
\vspace{-0.5cm}
   \input{echBWn3.tex}\\
  \end{tabular}
\caption{ECH and $\overline{\text{MS}}$ PT approximations to data at NLO and NNLO for $\langle B_W^n\rangle$, with n=1,2,3 from top to bottom.}
\label{ECH plot BW}
 \end{center}
\end{figure}

\begin{figure}
 \begin{center}
  \begin{tabular}{c}
\vspace{-0.5cm}
   \input{echBTn1.tex}\\
\vspace{-0.5cm}
   \input{echBTn2.tex}\\
\vspace{-0.5cm}
   \input{echBTn3.tex}\\
  \end{tabular}
\caption{ECH and $\overline{\text{MS}}$ PT approximations to data at NLO and NNLO for $\langle B_T^n\rangle$, with n=1,2,3 from top to bottom.}
\label{ECH plot BT}
 \end{center}
\end{figure}

\begin{figure}
 \begin{center}
  \begin{tabular}{c}
\vspace{-0.5cm}
   \input{echY3n1.tex}\\
\vspace{-0.5cm}
   \input{echY3n2.tex}\\
\vspace{-0.5cm}
   \input{echY3n3.tex}\\
  \end{tabular}
\caption{ECH and $\overline{\text{MS}}$ PT approximations to data at NLO and NNLO for $\langle Y_3^n\rangle$, with n=1,2,3 from top to bottom.}
\label{ECH plot Y3}
 \end{center}
\end{figure}

\begin{figure}
 \begin{center}
  \input{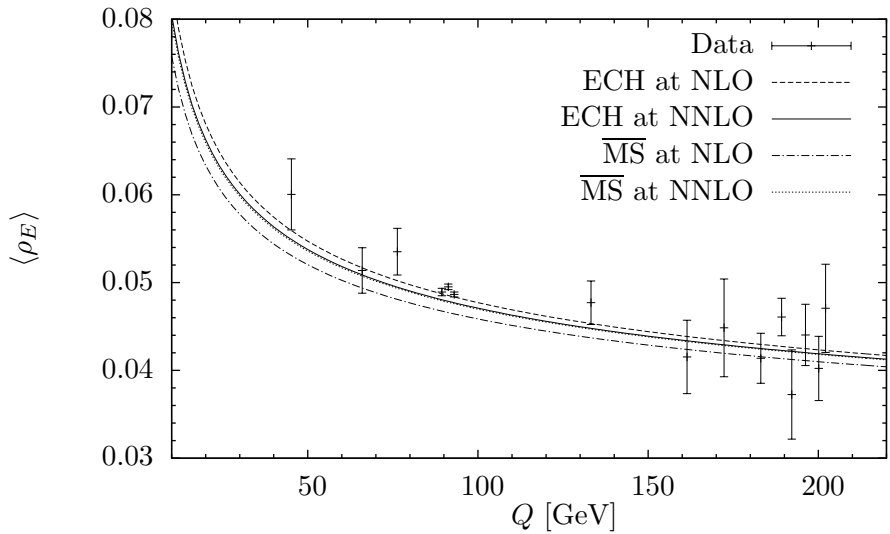}
\caption{ECH and $\overline{\text{MS}}$ PT approximations to data at NLO and NNLO for $\langle \rho_E\rangle$.}
\label{ECH plot rhoEn1}
 \end{center}
\end{figure}

The NLO and NNLO ECH approximations are shown against the data in Figs \ref{ECH plot T}-\ref{ECH plot rhoEn1}. For comparison the $\overline{\text{MS}}$ PT predictions are also shown. These are obtained by running the coupling from the fixed $\alpha_s(M_z)$ over a range of energies using \eqref{alpha=} and \eqref{alpha2=}, at NLO or NNLO as appropriate. The data are then approximated using \eqref{pt exp y}.

\subsection{Commentary on the results}\label{Commentary1}

Since the various tables and figures contain rather a lot of information it will be useful at this stage to distill some key conclusions from them. The values of $\alpha_s(M_z)$ extracted in Section \ref{Finding alpha} agree well with those found by the DELPHI collaboration in~\cite{DELPHI}. They are well grouped, as can be seen in Fig. \ref{scatter alpha} (top left panel); the values for $B_W$ and $Y_3$ are slight outliers. The average for NLO ECH agrees very well with the N$^3$LO Z-decay calculation value~\cite{Baikov}. The average for the $\overline{\text{MS}}$ perturbation theory ($\overline{\text{MS}}$ PT) calculation at NLO (bottom left panel) sits further from the N$^3$LO calculation, and the scatter of the central values is larger than for ECH at NLO. The numerical values of $\alpha_s(M_Z)$ are given in Tables \ref{table1} and \ref{table MSbar}.

We also demonstrated, in Section \ref{ECH plots}, how the ECH method can be inverted to give an approximation to data. Figs. \ref{ECH plot T}-\ref{ECH plot rhoEn1} show that the data for the means are best described by the NLO ECH curves, which lie closer to data than both the NLO and NNLO $\overline{\text{MS}}$ PT curves. The exceptions to this are $\langle B_W\rangle$, for which all the perturbative techniques give a very similar description of the data, and $\langle Y_3 \rangle$, where the $\overline{\text{MS}}$ PT at NLO curve describes the data best.

However the ECH method at NLO works less well for higher moments. This is particularly the case for $\langle(1-T)^n\rangle$, $\langle C^n\rangle$ and $\langle B_T^n\rangle$ with $n=2,3$, as can be seen in the bottom two panels of Figs. \ref{ECH plot T}, \ref{ECH plot C} and \ref{ECH plot BT}. $\overline{\text{MS}}$ PT also does not do a good job at describing the higher moment data.

ECH at NNLO describes the data less well than ECH at NLO, as shown in Section \ref{ECH plots}. The exception to this observation is $\langle Y_3\rangle$, in Fig. \ref{ECH plot Y3}, for which the NNLO ECH curve lies closer to the data than that for NLO ECH. Also, there is very little difference between any of the curves for $\langle B_W\rangle$. The NNLO ECH curves tend to give a description of the means that is similar to or better than the $\overline{\text{MS}}$ PT curves. For the higher moments ECH at NNLO is again often very similar to the $\overline{\text{MS}}$ PT curves, although it sometimes fails to describe the behaviour of the low energy data as well as $\overline{\text{MS}}$ PT does (e.g. $\langle C^2\rangle$ in Fig. \ref{ECH plot C}).

In the top right panel of Fig. \ref{scatter alpha}, we see that the values of $\alpha_s(M_z)$ are still fairly well grouped when extracted using ECH at NNLO, with $\langle Y_3\rangle$ remaining an outlier, but that the average has moved away from the N$^3$LO result. $\overline{\text{MS}}$ PT at NNLO, shown in the bottom right panel, is seen to have improved from the $\overline{\text{MS}}$ PT at NLO case: there are smaller errors due to a reduced scale dependence at higher order, the grouping is improved and the average is closer to the N$^3$LO calculation. In the higher moment plots (Figs. \ref{scatter alpha n2} and \ref{scatter alpha n3}) the grouping in the ECH method gets worse whereas the improvement seen going from NLO to NNLO in $\overline{\text{MS}}$ PT continues to hold.

These results confirm the success of NLO ECH in providing an accurate description of event shape means, as was noted in the analysis of~\cite{DELPHI}. There is generally no need to add a non-perturbative power correction. However, for $\overline{\text{MS}}$ PT at NLO additional power corrections are often necessary; these are often obtained using a dispersive model of power corrections~\cite{Luisoni, DokshitzerWebber, SalamWicke}. Whether these power corrections are really required, given the success of NLO ECH, is an issue that has been actively debated, in particular at a FRIF workshop held in Paris in 2006~\cite{FRIF}. In Section \ref{Commentary2}, we will discuss the results of fits using this dispersive model. A new feature of these fits is that the ECH method is used in conjunction with the dispersive power corrections.

Before exploring power correction models, we attempt to explain why the ECH method at NNLO does not provide a good description of the data, even for event shape means. We also try to motivate why NLO ECH does not give an accurate description of the higher moments. We do this by studying the relative sizes of terms in the asymptotic series expansion of $\rho(\mathcal{R})$, and examining the size of $\Lambda_{\mathcal{R}}$ for the variables, as shown in the next section.

\subsection{Terms in the expansion of $\rho(\mathcal{R})$ and the size of $\Lambda_{\mathcal{R}}$ values}\label{rho(R)}

In order to attempt to explain the results in Section \ref{Results}, we now look at the expansions $\rho(\mathcal{R})$. From Section \ref{Theory} we have:
\begin{equation}
 \rho(\mathcal{R}) = \frac{d\mathcal{R}(Q)}{d\ln Q} =
-b\mathcal{R}^2(1+c\mathcal{R}+\rho_2\mathcal{R}^2+\dots)\;.
\end{equation}
Perturbative expansions in QCD and other quantum field theories are expected to be asymptotic series, with an $n!$ growth of coefficients in $n^{\rm{th}}$ order perturbation theory. This divergent large-order behaviour arises from renormalon diagrams involving chains of vacuum polarization bubbles (for a review see~\cite{Beneke}). The large-order behaviour of the $\rho(\mathcal{R})$ function for the inclusive QCD $R_{e^+e^-}$ ratio has been investigated using renormalon techniques in~\cite{Tonge}, and has been shown to be an asymptotic series. 

A series that is asymptotic to a function, $F(a)$, has the following property (see~\cite{BenderOrszag} for a discussion of asymptotic series):
\begin{equation}\label{asymp series}
 F(a)\approx\sum_{n=0}^\infty f_n a^n \qquad \text{if} \qquad
\left|F(a)-\sum_{n=0}^N f_n a^n\right| \leq \left|f_{N+1} a^{N+1}\right|\;,
\end{equation}
where $\approx$ indicates that the series $\sum f_n a^n$ is asymptotic to the function $F(a)$. The crucial feature of \eqref{asymp series} is that the error associated with terminating the series at a particular point (after $N$ terms) is bounded by the size of the first term neglected (the $(N+1)^{\text{th}}$ term in this case). To get the best approximation to $F(a)$ we therefore want to terminate the series just before the term of smallest absolute magnitude.  This is a very important property of asymptotic series; although they are divergent, with zero radius of convergence, one can still obtain accurate approximations by truncating them after a suitable number of terms.

In Table \ref{rhoterms} we show the relative sizes of the known terms in $\frac{\rho(\mathcal{R})}{-b\mathcal{R}^2}$ for the event shape moments: $c\mathcal{R}$, the NLO term, and $\rho_2\mathcal{R}^2$, the NNLO term. These numbers are calculated using data from the $M_z$ peak. The leading order (LO) term in $\frac{\rho(\mathcal{R})}{-b\mathcal{R}^2}$, 1, is universal. The last column indicates the order at which the series should be terminated based on the relative magnitudes of the currently known terms. Note that, where $\rho_2\mathcal{R}^2$ is the smallest term, it is not possible to say definitively whether the series should be terminated at NLO or NNLO without knowing the size of the N$^3$LO term.

{\renewcommand{\arraystretch}{1.5}
\renewcommand{\tabcolsep}{0.3cm}
\begin{table}
\small
\begin{center}
 \begin{tabular}{>{\centering}p{1.8cm} >{\centering}p{0.8cm} >{\centering}p{1.6cm} >{\centering}p{1.6cm} >{\centering}p{3.6cm}}
  \hline
\hline
Observable & n & $c\mathcal{R}$ \, (NLO) & $\rho_2\mathcal{R}^2$ (NNLO) & Order at which series should be terminated
\tabularnewline
\hline
$\langle (1-T)^n\rangle$ & 1 & 0.078 & $-0.049$ & NLO
\tabularnewline
& 2 & 0.105 & $-0.323$ & LO
\tabularnewline
& 3 & 0.119 & $-0.575$ & LO
\tabularnewline
\hline
$\langle C^n\rangle$ & 1 & 0.076 & $-0.030$ & NLO
\tabularnewline
& 2 & 0.106 & $-0.376$ & LO
\tabularnewline
& 3 & 0.123 & $-0.912$ & LO
\tabularnewline
\hline
$\langle B_W^n\rangle$ & 1 & 0.045 & $\phantom{-}0.004$ & NLO
\tabularnewline
& 2 & 0.057 & $-0.031$ & NLO
\tabularnewline
& 3 & 0.055 & $-0.039$ & NLO
\tabularnewline
\hline
$\langle B_T^n\rangle$ & 1 & 0.066 & $-0.060$ & NLO
\tabularnewline
& 2 & 0.111 & $-1.803$ & LO
\tabularnewline
& 3 & 0.140 & $-5.715$ & LO
\tabularnewline
\hline
$\langle Y_3^n\rangle$ & 1 & 0.058 & $-0.042$ & NLO
\tabularnewline
& 2 & 0.060 & $-0.049$ & NLO
\tabularnewline
& 3 & 0.060 & $-0.053$ & NLO
\tabularnewline
\hline
$\langle \rho_E^n\rangle$ & 1 & 0.059 & $-0.015$ & NLO
\tabularnewline
\hline
\hline
 \end{tabular}
\caption{Size of terms in $\frac{\rho(\mathcal{R})}{-b\mathcal{R}^2}$}
\label{rhoterms}
\end{center}
\end{table}}

This information can help explain the results discussed in Section \ref{Commentary1}. For example, the expansions for the higher moments of $\langle(1-T)^n\rangle$, $C^n$ and $B_T^n$ all suggest that it is best to terminate the series at leading order. This indicates that using perturbation theory with this type of expansion, where the observable itself is used as the expansion parameter, is inappropriate for these event shape moments. As discussed previously, it can be seen that applying the ECH method to the higher moments of these observables does not work well.  However, in some instances (e.g. \mbox{$\langle (1-T)^2\rangle$} in Fig. \ref{ECH plot T}) the ECH method at NLO appears to be converging with the data at high energies. This implies that we are examining data below the perturbative regime of the method for these variables, but that if we go to sufficiently high energies the ECH method may be found to describe the data again. As $Q$ increases $\mathcal{R}(Q)$ decreases; for a suitably high energy the NNLO truncation will therefore represent the optimal number of terms, and will successfully describe the data. Of course this is of theoretical rather than practical interest, since we do not have data at indefinitely large energies.

The expansions for all other variables in Table \ref{rhoterms} suggest that it is preferable to terminate at NLO. As discussed in Section \ref{Commentary1}, it appears that the ECH method generally works better at NLO than at NNLO, although a firm conclusion cannot be made in these cases since we have no information on the size of the N$^3$LO term. We can also see that the higher moments of $B_W$ and $Y_3$ do not display the same undesirable behaviour as higher moments of the other variables: see the bottom two panels of Figs. \ref{ECH plot BW} and \ref{ECH plot Y3}.

We also examine the size of $\Lambda_{\mathcal{R}}$ for each variable. The values of $\Lambda_{\mathcal{R}}$, shown in Table \ref{lambdaRtable}, are calculated from a fixed value of $\alpha_s(M_Z)=0.1190$ and \eqref{Lambda twiddle}, which when inverted gives $\Lambda_{\mathcal{R}}=e^{-r/b} \,\tilde{\Lambda}_{\overline{\text{MS}}}$. From \eqref{int2} we see that ${\mathcal{R}}(Q)$ has a Landau pole when $Q=\Lambda_{\mathcal{R}}$, and will formally diverge there~\cite{QCDreview}. From Table \ref{lambdaRtable}, the $\Lambda_{\mathcal{R}}$ values for the n=1 moments are a few GeV. Therefore the data we are using are at energies considerably higher than the Landau pole and the ECH method should work well. For the higher moments, the values of $\Lambda_{\mathcal{R}}$ increase. We are therefore analysing data in a region of $Q$ which contains the Landau pole, and the perturbative ECH method will not function until we have data at much larger energies. This is the same conclusion that we reached by studying the magnitudes of terms in the $\rho(\mathcal{R})$ series. The $\langle B_W^n\rangle$ and $\langle Y_3^n\rangle$ moments for $n=2,3$ are an exception, with small values of $\Lambda_{\mathcal{R}}$. As stated above, the ECH method performs better for these higher moments compared with the other observables.

{\renewcommand{\arraystretch}{1.5}
\renewcommand{\tabcolsep}{0.3cm}
\begin{table}
 \begin{center}
  \begin{tabular}{cccccccc}
\hline
\hline
n & & $\langle (1-T)^n\rangle$ & $\langle C^n\rangle$ & $\langle B_W^n\rangle$ & $\langle B_T^n\rangle$ & $\langle Y_3^n\rangle$ & $\langle \rho_E^n\rangle$\\
\hline
1 & & 3 & 3 & 0.1 & 2 & 1 & 1 \\
2 & & 14 & 15 & 1.1 & 58 & 2 & -\\
3 & & 27 & 34 & 1.2 & 384 & 1 & -\\
\hline
\hline
  \end{tabular}
\caption{Values of $\Lambda_{\mathcal{R}}$ in GeV \label{lambdaRtable}}
 \end{center}
\end{table}}

\subsection{Dispersive power corrections}\label{pcs}

In this section we use the ECH method in combination with a dispersive model for power corrections~\cite{Luisoni, DokshitzerWebber, SalamWicke}. Hadronisation corrections for event shape moments are expected to be additive~\cite{Luisoni, Glover}:
\begin{equation}\label{pt+np}
 \langle y^n \rangle=\langle y^n \rangle_{pt} + \langle y^n \rangle_{np}\;,
\end{equation}
where $\langle y^n \rangle_{pt}$ is the perturbative part of the event shape moment and $\langle y^n \rangle_{np}$ is the non-perturbative part.

The dispersive model for power corrections accounts for non-perturbative behaviour at low energies by replacing the strong coupling constant with an effective coupling, $\alpha_{eff}$, below an IR cutoff scale, $\mu_I$. This is done in such a way that the integral of the effective coupling up to $\mu_I$ is finite:
\begin{equation}
 \frac{1}{\mu_I}\int_0^{\mu_I} dQ\,\alpha_{\text{eff}}(Q^2)=\alpha_0(\mu_I)\;.
\end{equation}
This leads to the following non-perturbative contribution to the event shape means:
\begin{equation}\label{np forms}
 \langle y\, \rangle_{np} = a_y P\;,
\end{equation}
where the $a_y$ are numerical factors that depend on the event shape in question~\cite{Luisoni}. Their values are given in Table \ref{ay=}. Note that since $a\,_{\text{\tiny{\textit{Y}}}_3}\!=0$ there are no dispersive power corrections for $\langle Y_3^n\rangle$.

The form of $P$ at NNLO is:
\begin{align}\label{P=}
 P=\frac{4C_F}{\pi^2}\mathcal{M}\bigg(\alpha_0-\bigg[&\alpha_s(\mu_R)+\frac{b}{\pi}\left(1+\ln\frac{\mu_R}{\mu_I} +\frac{K}{2b}\right)\alpha_s^2(\mu_R)
\nonumber\\
 &+\bigg(4bc\left(1+\ln\frac{\mu_R}{\mu_I}+\frac{L}{4bc}\right)+8b^2\left(1+\ln\frac{\mu_R}{\mu_I}+\frac{K}{2b}\right) \nonumber\\
&\qquad+4b^2\ln\frac{\mu_R}{\mu_I}\left(\ln\frac{\mu_R}{\mu_I}+\frac{K}{b}\right)\bigg)\frac{\alpha_s^3(\mu_R)}{4\pi^2}\bigg]\bigg)\frac{\mu_I}{Q}\;,
\end{align}
where $C_F=\frac{N^2-1}{2N}$ (with $N=3$ being the number of colours), and:
\begin{align*}
 K=&\left(\frac{67}{18}-\frac{\pi^2}{6}\right)C_A-\frac{5}{9}N_F\;,\\
 L=&\;C_A^2\left(\frac{245}{24}-\frac{67}{9}\frac{\pi^2}{6}+\frac{11}{6}\zeta_3+\frac{11}{5}\Big(\frac{\pi^2}{6}\Big)^2\right)+C_FN_F\left(-\frac{55}{24}+2\zeta_3\right)\\
&\qquad\qquad+C_AN_F\left(-\frac{209}{108}+\frac{10}{9}\frac{\pi^2}{6}-\frac{7}{3}\zeta_3\right)-\frac{1}{27}N_F^2\;,
\end{align*}
where $C_A=N$ and $N_F$ is the number of quark flavours.  $\mathcal{M}$ is the Milan factor, a two-loop enhancement factor~\cite{Milan}. $\mathcal{M}$ is universal for observables with this type of $1/Q$ power correction. Given the energy range of the data used in this paper, the number of active quark flavours is 5. The form of $P$ at NLO ($P^{(NLO)}$) is the same as $P$ in \eqref{P=} except that the $\alpha_s^3$ term in the square brackets is omitted.

For $B_W$ and $B_T$ there are additional corrections to $P$. At NLO we get:
\begin{align}
 P_{B_W}^{(NLO)}&=P^{(NLO)}\left(\frac{\pi}{\sqrt{8C_F\hat{\alpha}_s\left(1+\frac{K\hat{\alpha}_s}{2\pi}\right)}}+\frac{3}{4}-\frac{b}{6C_F}+\eta_0\right)\;,\\
 P_{B_T}^{(NLO)}&=P^{(NLO)}\left(\frac{\pi}{\sqrt{4C_F\hat{\alpha}_s\left(1+\frac{K\hat{\alpha}_s}{2\pi}\right)}}+\frac{3}{4}-\frac{b}{3C_F}+\eta_0\right)\;,
\end{align}
where $\hat{\alpha}_s(\mu_R)\equiv\alpha_s(e^{-3/4}\mu_R)$ and $\eta_0=-0.6137$. The corrections to $P$ for $B_W$ and $B_T$ at NNLO have not been explicitly calculated yet but the potentially dominant terms are approximated by~\cite{Luisoni}:
\begin{align}
 P_{B_W}&=P\left(\frac{\pi}{\sqrt{8C_F\hat{\alpha}_s\left(1+\frac{K\hat{\alpha}_s}{2\pi}+\frac{L\hat{\alpha}_s^2}{4\pi^2}\right)}}+\frac{3}{4}-\frac{b}{6C_F}+\eta_0\right)\;,\\
 P_{B_T}&=P\left(\frac{\pi}{\sqrt{4C_F\hat{\alpha}_s\left(1+\frac{K\hat{\alpha}_s}{2\pi}+\frac{L\hat{\alpha}_s^2}{4\pi^2}\right)}}+\frac{3}{4}-\frac{b}{3C_F}+\eta_0\right)\;.
\end{align}

In this section we use the forms in \eqref{pt+np} and \eqref{np forms} to predict the form of the data and then perform simultaneous fits for $\alpha_s(M_z)$ and $\alpha_0$. The parameter $\alpha_0$ is expected to be universal. For $\langle y\rangle_{pt}$ we use the ECH prediction at NLO and NNLO (see section \ref{ECH plots}). In order to facilitate inversion of \eqref{Lambda=} at NNLO we use Pad\'{e} Approximant (PA) methods~\cite{PApaper,
Burrows, Burrows2, SamuelEllisKarliner, EllisGardiKarlinerSamuel}. We do this by writing \eqref{dR/dlnx} at NNLO as $x^2$ multiplied by a [1/1] PA:
\begin{equation}
 \rho(x)=-b x^2 \left(\frac{1+(c-\frac{\rho_2}{c})x}{1-(\frac{\rho_2}{c}) x} + \mathcal{O}(x^3)\right)\;.
\end{equation}
We can then integrate up and invert analytically to get:
\begin{equation}\label{R_NNLO}
\displaystyle
 \mathcal{R}_{\text{NNLO}}(Q)=\frac{-1}{c\left[1-\frac{\rho_2}{c^2}+W\left(-\exp\left[-\left(1-\frac{\rho_2}{c^2}+\frac{b}{c}\ln\frac{Q}{\Lambda_{\mathcal{R}}}\right)\right]\right)\right]}\;.
\end{equation}
In the expressions for the power correction term $\mu_R$ is set to \mbox{$\mu_{\text{ECH}}=e^{-r/b}\,Q$} (see the discussion below \eqref{R(Q)}) when the ECH perturbative part is used. 

The IR cutoff should be chosen such that $\Lambda_{\text{QCD}} \ll \mu_I \ll Q$. We therefore take $\mu_I$ to be 2 GeV. Since the fits for $\alpha_s$ and $\alpha_0$ have a dependency on the choice of cutoff, we vary $\mu_I$ between 1 and 3 GeV to assess the extent of the dependency. The variation in the fits is used to assign a theoretical uncertainty. This is in keeping with the method used in~\cite{Luisoni}. The Milan factor is known at two loop order to be $\mathcal{M}=1.49\pm20$\%, so $\mathcal{M}$ is also varied around this central value. The central, upper and lower values of $\mu_I$ and $\mathcal{M}$ are shown in Table \ref{variation}. When the effect of the upward and downward variation result in uneven differences the larger of the two is taken to contribute towards the theoretical uncertainty. The individual differences from the variation of $\mu_I$ and $\mathcal{M}$ are added together in quadrature to give the theoretical uncertainty.

{\renewcommand{\arraystretch}{1.5}
\renewcommand{\tabcolsep}{0.2cm}
\begin{table}
\small
\begin{center}
 \begin{tabular}{cccccccc}
\hline
\hline
Event shape & \hspace{0.25cm} & $1-T$ & $C$ & $B_W$ & $B_T$ & $Y_3$ & $\rho_E$\\
\hline
$a_y$ & & 2 & 3$\pi$ & 1/2 & 1 & 0 & 1\\
\hline
\hline
 \end{tabular}
\caption{The $a_y$ coefficients for different event shapes \label{ay=}}
\end{center}
\end{table}}

\begin{table}
\small
 \begin{center}
  \begin{tabular}{ccccccc}
\hline
\hline
  & \hspace{0.25cm} & Central value & \hspace{0.1cm} & Upper value & \hspace{0.1cm} & Lower value\\
\hline
  $\mu_I$ [GeV] & & 2 & & 3 & & 1 \\
  $\mathcal{M}$ & & 1.49 & & 1.788 ($+20$\%) & & 1.192 ($-20$\%)\\
  $\mu_R$ [GeV] & & $Q$ & & $2Q$ & & $Q/2$\\
\hline
\hline
  \end{tabular}
\caption{Variation of $\mu_I$, $\mathcal{M}$ and $\mu_R$ \label{variation}}
 \end{center}
\end{table}

Since it was found in the previous sections that generally higher moments work less well than the $n=1$ moments, we confine ourselves to analysing the means in this part of the paper. The results of the dual fits for means at NLO and NNLO are shown in Table \ref{fit results}. The first uncertainty is the combined experimental and perturbative coefficient uncertainty, and the second is the theoretical uncertainty (estimated by varying $\mu_I$ and $\mathcal{M}$). These results are plotted in Fig. \ref{scatter alpha0}. We also perform fits using the $\overline{\text{MS}}$ PT prediction at NLO and NNLO. These results are shown in Table \ref{MSbar+pc} and Fig. \ref{scatter alpha0 MSbar}. In these cases $\mu_R$ is varied (around a central value of $\mu_R=Q$), as well as $\mu_I$ and $\mathcal{M}$, to find the theoretical uncertainty (see Table \ref{variation}).

{\renewcommand{\arraystretch}{1.5}
\renewcommand{\tabcolsep}{0.3cm}
\begin{table}
\small
 \begin{center}
  \begin{tabular}{cccc}
   \hline
\hline
Observable & $\alpha_s(M_z)$ & $\alpha_0$ & $\chi^2$/dof\\
\hline
\multicolumn{4}{c}{NLO}\\
\hline
$\langle 1-T\rangle$ & $0.1144\pm0.0007\pm0.0028$ & $0.436\pm0.011\pm0.006$ & 52.6/46\\
$\langle C\rangle$ & $0.1133\pm0.0006\pm0.0031$ & $0.423\pm0.006\pm0.010$ & 58.5/40\\
$\langle B_W\rangle$ & $0.1192\pm0.0006\pm0.0014$ & $0.307\pm0.014\pm0.011$ & 41.4/46\\
$\langle B_T\rangle$ & $0.1188\pm0.0010\pm0.0038$ & $0.312\pm0.027\pm0.053$ & 59.8/46\\
$\langle \rho_E\rangle$ & $0.1163\pm0.0055\pm0.0010$ & $0.342\pm0.136\pm0.012$ & 11.6/13\\
\hline
\hline
\multicolumn{4}{c}{NNLO}\\
\hline
$\langle 1-T\rangle$ & $0.1142\pm0.0006\pm0.0017$ & $0.584\pm0.010\pm0.027$ & 58.3/46\\
$\langle C\rangle$ & $0.1130\pm0.0005\pm0.0019$ & $0.539\pm0.008\pm0.026$ & 67.0/40\\
$\langle B_W\rangle$ & $0.1184\pm0.0007\pm0.0014$ & $0.392\pm0.014\pm0.013$ & 43.1/46\\
$\langle B_T\rangle$ & $0.1184\pm0.0008\pm0.0028$ & $0.537\pm0.020\pm0.008$ & 67.2/46\\
$\langle \rho_E\rangle$ & $0.1168\pm0.0054\pm0.0010$ & $0.449\pm0.114\pm0.017$ & 11.6/13\\
\hline
\hline
  \end{tabular}
\caption{Fits for $\alpha_s(M_z)$ and $\alpha_0$ using ECH and dispersive power corrections at NLO and NNLO.}
\label{fit results}
 \vspace{2cm}
  \begin{tabular}{cccc}
   \hline
\hline
Observable & $\alpha_s(M_z)$ & $\alpha_0$ & $\chi^2$/dof\\
\hline
\multicolumn{4}{c}{NLO}\\
\hline
$\langle 1-T\rangle$ & $0.1302\pm0.0010\pm0.0088$ & $0.382\pm0.012\pm0.012$ & 49.7/46\\
$\langle C\rangle$ & $0.1274\pm0.0008\pm0.0085$ & $0.364\pm0.007\pm0.013$ & 48.5/40\\
$\langle B_W\rangle$ & $0.1191\pm0.0007\pm0.0021$ & $ 0.346\pm0.015\pm0.126$ & 42.7/46\\
$\langle B_T\rangle$ & $0.1252\pm0.0008\pm0.0060$ & $0.312\pm0.013\pm0.005$ & 64.3/46\\
$\langle \rho_E\rangle$ & $0.1200\pm0.0060\pm0.0042$ & $0.323\pm0.138\pm0.018$ & 11.5/13\\
\hline
\hline
\multicolumn{4}{c}{NNLO}\\
\hline
$\langle 1-T\rangle$ & $0.1216\pm0.0009\pm0.0038$ & $0.427\pm0.011\pm0.020$ & 49.5/46\\
$\langle C\rangle$ & $0.1197\pm0.0007\pm0.0039$ & $0.412\pm0.007\pm0.022$ & 47.9/40\\
$\langle B_W\rangle$ & $0.1179\pm0.0007\pm0.0017$ & $0.429\pm0.015\pm0.045$ & 44.3/46\\
$\langle B_T\rangle$ & $0.1212\pm0.0008\pm0.0029$ & $0.385\pm0.012\pm0.038$ & 65.1/46\\
$\langle \rho_E\rangle$ & $0.1178\pm0.0056\pm0.0014$ & $0.389\pm0.127\pm0.033$ & 11.6/13\\
\hline
\hline
  \end{tabular}
\caption{Fits for $\alpha_s(M_z)$ and $\alpha_0$ using $\overline{\text{MS}}$ PT and dispersive power corrections at NLO and NNLO.}
\label{MSbar+pc}
 \end{center}
\end{table}}

\begin{figure}
 \begin{center}
  \begin{tabular}{cc}
  \input{scatter_alphas_NLO.tex} & \hspace{-1.75cm}
\input{scatter_alpha0_NLO.tex}\\
  \input{scatter_alphas_NNLO.tex} & \hspace{-1.75cm}
\input{scatter_alpha0_NNLO.tex}
  \end{tabular}
\caption{Scatter plots of $\alpha_s(M_z)$ and $\alpha_0$ from fits of ECH and dispersive power corrections for the means at NLO (top two panels) and NNLO (bottom two panels). The dotted lines on the $\alpha_s(M_z)$ plots show the value of the coupling obtained from N$^3$LO calculations\protect{~\cite{Baikov}}. The solid lines show the unweighted averages.}
\label{scatter alpha0}
 \end{center}
\end{figure}

\begin{figure}
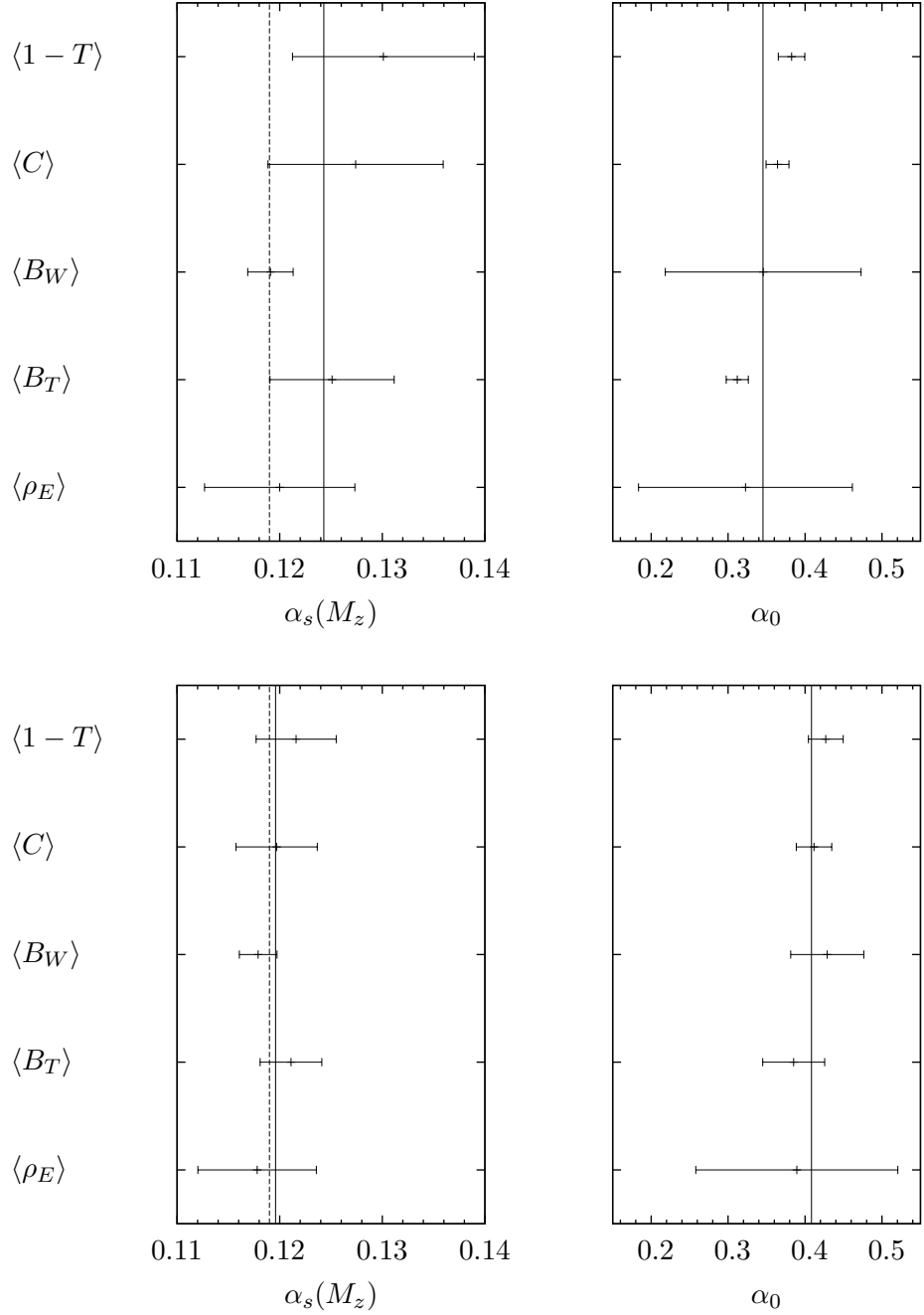

 \begin{center}
  \begin{tabular}{cc}
  \input{scatter_alphas_NLO_MSbar.tex} & \hspace{-1.75cm}
\input{scatter_alpha0_NLO_MSbar.tex}\\
   \input{scatter_alphas_NNLO_MSbar.tex} & \hspace{-1.75cm}
\input{scatter_alpha0_NNLO_MSbar.tex}
  \end{tabular}
\caption{Scatter plots of $\alpha_s(M_z)$ and $\alpha_0$ from fits of $\overline{\text{MS}}$ PT and dispersive power corrections at NLO (top two panels) and NNLO (bottom two panels). The solid and dotted lines are as in Fig. \ref{scatter alpha0}.}
\label{scatter alpha0 MSbar}
 \end{center}
\end{figure}

In analogy to Section \ref{ECH plots}, we also perform fits for $\alpha_0$ while keeping $\alpha_s(M_z)$ fixed at 0.1190. This is done for both ECH and standard $\overline{\text{MS}}$ PT at NLO and NNLO, and the results are shown in Tables \ref{single fits ECH}-\ref{single fits MSbar} and Figs. \ref{scatter single fit ECH}-\ref{scatter single fit MSbar}.

\begin{table}
\small
 \begin{center}
  \begin{tabular}{ccccc}
\hline
\hline
Observable & \multicolumn{4}{c}{$\alpha_0$}\\
\hline
   & NLO & $\chi^2$/dof & NNLO & $\chi^2$/dof\\
\hline
$\langle 1-T\rangle$ & $0.377\pm0.007\pm0.003$ & 101.6/47 & $0.557\pm0.013\pm0.019$ & 166.3/47\\
$\langle C\rangle$ & $0.375\pm0.007\pm0.004$ & 209.6/41 & $0.524\pm0.013\pm0.013$ & 374.8/41\\
$\langle B_W\rangle$ & $0.317\pm0.005\pm0.028$ & 42.0/47 & $0.381\pm0.014\pm0.025$ & 43.9/47\\
$\langle B_T\rangle$ & $0.314\pm0.004\pm0.007$ & 59.8/47 & $0.526\pm0.017\pm0.036$ & 68.0/47\\
$\langle \rho_E\rangle$ & $0.282\pm0.010\pm0.008$ & 11.8/14 & $0.403\pm0.015\pm0.016$ & 11.8/14\\
\hline
\hline
  \end{tabular}
\caption{Single fits for $\alpha_0$, with $\alpha_s(M_z)=0.1190$ held fixed, for ECH and dispersive power corrections at NLO and NNLO.}
\label{single fits ECH}
\vspace{2cm}
  \begin{tabular}{ccccc}
\hline
\hline
Observable & \multicolumn{4}{c}{$\alpha_0$}\\
\hline
   & NLO & $\chi^2$/dof & NNLO & $\chi^2$/dof\\
\hline
$\langle 1-T\rangle$ & $0.504\pm0.010\pm0.088$ & 185.3/47 & $0.457\pm0.006\pm0.043$ & 59.8/47\\
$\langle C\rangle$ & $0.433\pm0.007\pm0.063$ & 198.8/41 & $0.418\pm0.005\pm0.028$ & 49.5/41\\
$\langle B_W\rangle$ & $0.353\pm0.005\pm0.154$ & 43.0/47 & $0.409\pm0.016\pm0.041$ & 46.9/47\\
$\langle B_T\rangle$ & $0.413\pm0.004\pm0.071$ & 167.2/47 & $0.419\pm0.010\pm0.039$ & 79.7/47\\
$\langle \rho_E\rangle$ & $0.352\pm0.010\pm0.088$ & 11.6/14 & $0.363\pm0.012\pm0.025$ & 11.6/14\\
\hline
\hline
  \end{tabular}
\caption{Single fits for $\alpha_0$, with $\alpha_s(M_z)=0.1190$ held fixed, for $\overline{\text{MS}}$ PT and dispersive power corrections at NLO and NNLO.}
\label{single fits MSbar}
 \end{center}
\end{table}

\begin{figure}
 \begin{center}
  \begin{tabular}{cc}
   \input{scatter_single_alpha0_NLO.tex} & \hspace{-1.75cm}
\input{scatter_single_alpha0_NNLO.tex}
  \end{tabular}
\caption{Scatter plots for single $\alpha_0$ fits for ECH and dispersive power corrections at NLO and NNLO.}
\label{scatter single fit ECH}

\vspace{0.75cm}
  \begin{tabular}{cc}
   \input{scatter_single_alpha0_NLO_MSbar.tex} & \hspace{-1.75cm}
\input{scatter_single_alpha0_NNLO_MSbar.tex}
  \end{tabular}
\caption{Scatter plots for single $\alpha_0$ fits for $\overline{\text{MS}}$ PT and dispersive power corrections at NLO and NNLO. The solid lines are unweighted averages.}
\label{scatter single fit MSbar}
 \end{center}
\end{figure}

\subsection{Simple power corrections}\label{simple pcs}

For comparison we now use a second model for non-perturbative corrections: a simple power correction~\cite{DELPHI, CampbellGloverMaxwell}. For the ECH method this is done by adding a $c_1$/Q power correction (with $c_1$ a constant) to the perturbative expansion of $\mathcal{R}$ given in \eqref{R=}. This results in an altered form for the $\rho(\mathcal{R})$ function as follows:
\begin{equation}
 \rho(\mathcal{R})=-b\mathcal{R}^2(1+c\mathcal{R}+\rho_2
\mathcal{R}^2+\dots)+\kappa_0 \mathcal{R}^{-c/b} e^{-1/(b\mathcal{R})}\;.
\end{equation}
This modified form of $\rho({R})$ is then substituted into \eqref{int2}, with the perturbative part truncated at NLO or NNLO, and a fit to data is performed to extract $\Lambda_\mathcal{R}$ (and hence $\alpha_s(M_z)$) and $\kappa_0$.

A corresponding power correction can be added to a NLO or NNLO $\overline{\text{MS}}$ perturbative part using:
\begin{align}
 \langle y^n\rangle =
\left(\frac{\alpha_s(\mu_R)}{2\pi}\right)\overline{\mathcal{A}}_{y,n}
+\left(\frac{\alpha_s(\mu_R)}{2\pi}\right)^2\overline{\mathcal{B}}_{y,n}&
+\left(\frac{\alpha_s(\mu_R)}{2\pi}\right)^3\overline{\mathcal{C}}_{y,n}+\dots
\nonumber\\
 &-\kappa_0 e^{r/b} \left(\frac{b}{2}\right)^{\frac{c}{b}}
\frac{\Lambda_{\overline{\text{MS}}}}{Q}\;.
\end{align}
In this model it is not expected that the fits will yield a universal value of $\kappa_0$. Each particular observable will have a different value of $\kappa_0$, unlike the dispersive model where the power correction parameter $\alpha_0$ is expected to be universal. The results are shown in Tables \ref{simple pc fit results}-\ref{simple pc fit results MSbar} and Figs. \ref{scatter kappa0}-\ref{scatter kappa0 MSbar}.

{\renewcommand{\arraystretch}{1.5}
\renewcommand{\tabcolsep}{0.4cm}
\begin{table}
\small
 \begin{center}
\vspace{-1.5cm}
  \begin{tabular}{cccc}
\hline
\hline
Observable & $\alpha_s(M_z)$ & $\kappa_0$ & $\chi^2$/dof\\
\hline
\multicolumn{4}{c}{NLO}\\
\hline
$\langle 1-T \rangle$ & $0.1230\pm0.0008$ & $-0.039\pm0.008$ & 51.3/46\\
$\langle C\rangle$ & $0.1230\pm0.0006$ & $-0.036\pm0.006$ & 43.7/40\\
$\langle B_W\rangle$ & $0.1225\pm0.0005$ & $\phantom{-}0.084\pm0.038$ & 34.2/46\\
$\langle B_T\rangle$ & $0.1225\pm0.0004$ & $-0.054\pm0.008$ & 60.7/46\\
$\langle Y_3\rangle$ & $0.1222\pm0.0021$ & $-0.237\pm0.045$ & 31.2/31\\
$\langle \rho_E\rangle$ & $0.1184\pm0.0052$ & $\phantom{-}0.047\pm0.181$ & 11.4/13\\
\hline
\hline
\multicolumn{4}{c}{NNLO}\\
\hline
$\langle 1-T\rangle$ & $0.1253\pm0.0009$ & $-0.016\pm0.009$ & 50.7/46\\
$\langle C\rangle$ & $0.1247\pm0.0008$ & $-0.022\pm0.008$ & 43.7/40\\
$\langle B_W\rangle$ & $0.1224\pm0.0006$ & $\phantom{-}0.077\pm0.038$ & 34.3/46\\
$\langle B_T\rangle$ & $0.1269\pm0.0010$ & $-0.020\pm0.010$ & 61.5/46\\
$\langle Y_3\rangle$ & $0.1261\pm0.0024$ & $-0.204\pm0.045$ & 31.2/31\\
$\langle \rho_E\rangle$ & $0.1196\pm0.0053$ & $\phantom{-}0.063\pm0.179$ & 11.4/13\\
\hline
\hline
  \end{tabular}
\caption{Fits for $\alpha_s(M_z)$ and $\kappa_0$ using ECH and simple power corrections at NLO and NNLO.}
\label{simple pc fit results}

\renewcommand{\arraystretch}{1.5}
\renewcommand{\tabcolsep}{0.3cm}
\vspace{1.4cm}
  \begin{tabular}{cccc}
\hline
\hline
Observable & $\alpha_s(M_z)$ & $\kappa_0$ & $\chi^2$/dof\\
\hline
\multicolumn{4}{c}{NLO}\\
\hline
$\langle 1-T \rangle$ & $0.1374\pm0.0013\pm0.0085$ & $-0.006\pm0.008\pm0.012$ & 51.5/46\\
$\langle C\rangle$ & $0.1357\pm0.0009\pm0.0079$ & $-0.026\pm0.028\pm0.057$ & 44.7/40\\
$\langle B_W\rangle$ & $0.1228\pm0.0005\pm0.0035$ & $-0.286\pm0.119\pm0.314$ & 34.6/46\\
$\langle B_T\rangle$ & $0.1305\pm0.0006\pm0.0063$ & $\phantom{-}0.042\pm0.016\pm0.051$ & 61.5/46\\
$\langle Y_3\rangle$ & $0.1260\pm0.0012\pm0.0054$ & $\phantom{-}0.060\pm0.008\pm0.022$ & 31.6/31\\
$\langle \rho_E\rangle$ & $0.1218\pm0.0062\pm0.0041$ & $-0.097\pm0.276\pm0.016$ & 11.4/13\\
\hline
\hline
\multicolumn{4}{c}{NNLO}\\
\hline
$\langle 1-T\rangle$ & $0.1279\pm0.0011\pm0.0025$ & $\phantom{-}0.002\pm0.012\pm0.020$ & 51.1/46\\
$\langle C\rangle$ & $0.1270\pm0.0008\pm0.0023$ & $\phantom{-}0.018\pm0.042\pm0.088$ & 44.1/40\\
$\langle B_W\rangle$ & $0.1223\pm0.0007\pm0.0003$ & $-0.238\pm0.129\pm0.122$ & 34.7/46\\
$\langle B_T\rangle$ & $0.1269\pm0.0007\pm0.0013$ & $\phantom{-}0.075\pm0.020\pm0.033$ & 61.1/46\\
$\langle Y_3\rangle$ & $0.1235\pm0.0012\pm0.0011$ & $\phantom{-}0.074\pm0.009\pm0.010$ & 31.6/31\\
$\langle \rho_E\rangle$ & $0.1199\pm0.0060\pm0.0008$ & $-0.089\pm0.307\pm0.016$ & 11.4/13\\
\hline
\hline
  \end{tabular}
\caption{Fits for $\alpha_s(M_z)$ and $\kappa_0$ using $\overline{\text{MS}}$ PT and simple power corrections at NLO and NNLO.}
\label{simple pc fit results MSbar}
 \end{center}
\end{table}}

\begin{figure}
 \begin{center}
  \begin{tabular}{cc}
  \input{scatter_simple_alphas_NLO.tex} & \hspace{-1.75cm}
\input{scatter_kappa0_NLO.tex}\\
  \input{scatter_simple_alphas_NNLO.tex} & \hspace{-1.75cm}
\input{scatter_kappa0_NNLO.tex}
  \end{tabular}
\caption{Scatter plots of $\alpha_s(M_z)$ and $\kappa_0$ from fits of ECH and simple power corrections for the means at NLO (top two panels) and NNLO (bottom two panels). The dotted lines on the $\alpha_s(M_z)$ plots show the value of the coupling obtained from N$^3$LO calculations\protect{~\cite{Baikov}}. The solid lines are unweighted averages.}
\label{scatter kappa0}
 \end{center}
\end{figure}

\begin{figure}
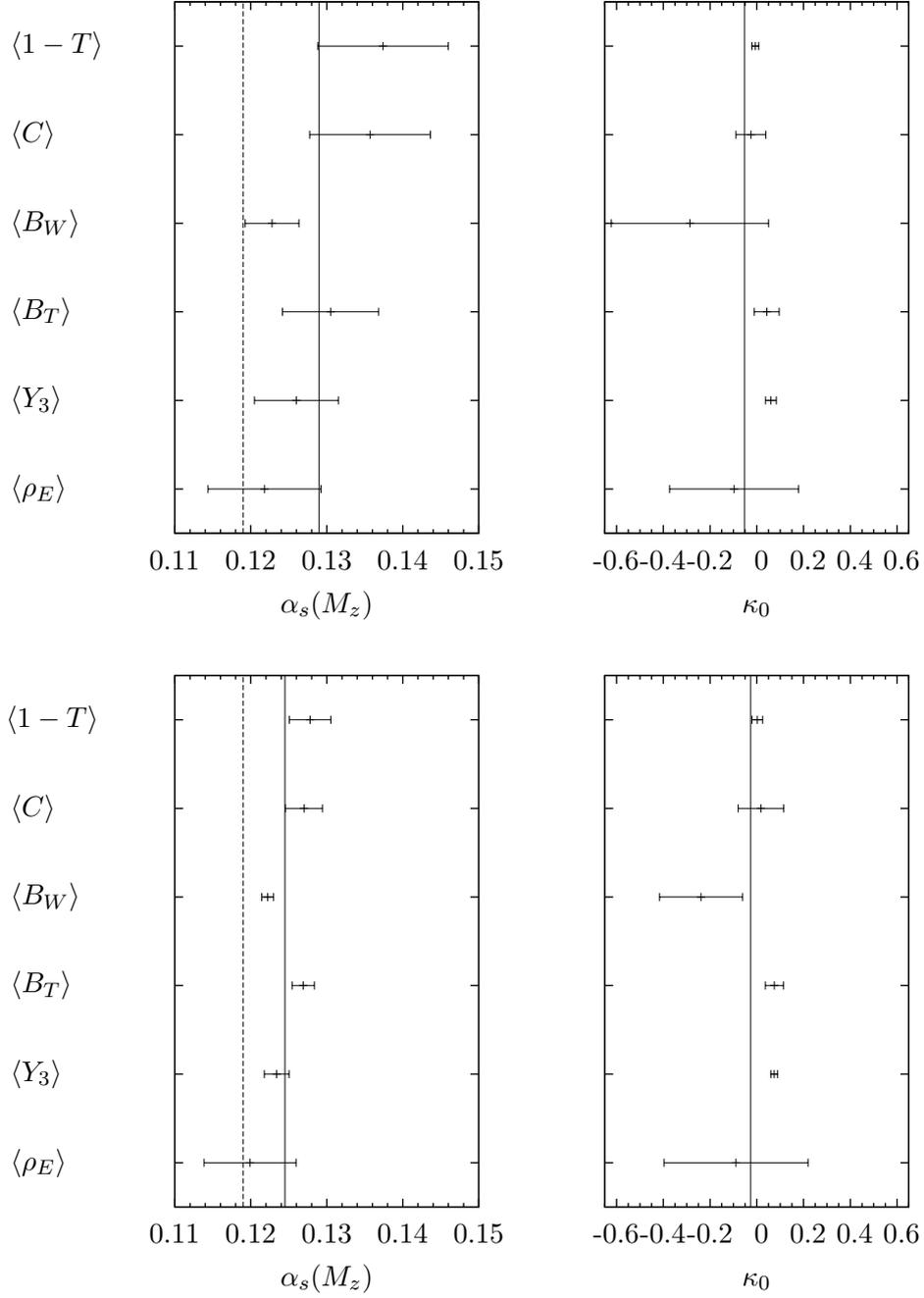

 \begin{center}
  \begin{tabular}{cc}
  \input{scatter_simple_alphas_NLO_MSbar.tex} & \hspace{-1.75cm} \input{scatter_kappa0_NLO_MSbar.tex}\\
  \input{scatter_simple_alphas_NNLO_MSbar.tex} & \hspace{-1.75cm} \input{scatter_kappa0_NNLO_MSbar.tex}
  \end{tabular}
\caption{Scatter plots of $\alpha_s(M_z)$ and $\kappa_0$ from fits of $\overline{\text{MS}}$ PT and simple power corrections for the means at NLO (top two panels) and NNLO (bottom two panels). The dotted and solid lines are as in Fig. \ref{scatter kappa0}}
\label{scatter kappa0 MSbar}
 \end{center}
\end{figure}

\subsection{Commentary on power correction fits}\label{Commentary2}

In Section \ref{pcs} we examine the effect of adding a dispersive power correction to the perturbative model. In Fig. \ref{scatter alpha0} it is seen that the values of $\alpha_s(M_z)$ and $\alpha_0$ extracted from $\langle B_W\rangle$, $\langle B_T\rangle$ and $\langle \rho_E\rangle$ using ECH plus a dispersive power correction at NLO agree well with each other. The results for \mbox{$\langle 1-T\rangle$} and $\langle C\rangle$ are somewhat discrepant. The values of $\alpha_s$ for the other three event shapes are not considerably changed from those extracted using pure ECH. This implies that only very small power corrections are required for ECH at NLO for these observables. At NNLO the extracted values of $\alpha_s(M_z)$ do not change substantially, but the fitted values of $\alpha_0$ become larger. The fits for $\langle 1-T\rangle$ and $\langle C\rangle$ are still not consistent with those from the other observables. The reduced $\chi^2$ values ($\chi^2$ divided by the number of degrees of freedom) for the fits generally get further from 1 when going from NLO to NNLO, indicating that the quality of the fits decreases.

For NLO $\overline{\text{MS}}$ PT plus a dispersive power correction, shown in Fig. \ref{scatter alpha0 MSbar}, the error bars are generally larger than for the corresponding ECH fits. This is because it is necessary to vary $\mu_R$, in addition to the Milan factor and the IR cutoff scale, when calculating the theoretical uncertainties. At NNLO these uncertainties get smaller, as expected, and furthermore the agreement between values of $\alpha_s(M_z)$ and $\alpha_0$ improves. There is no substantial change in the $\chi^2$ values between the NLO and NNLO fits for $\overline{\text{MS}}$ PT. The description of the data by $\overline{\text{MS}}$ PT plus dispersive power corrections is generally better than the corresponding ECH fits, as can be seen from the $\chi^2$ values in Tables \ref{scatter alpha0} and \ref{scatter alpha0 MSbar}.

In Fig. \ref{scatter single fit ECH}, when $\alpha_s(M_z)$ is held fixed and only $\alpha_0$ fitted for, good agreement is found for ECH at NLO. There is now less discrepancy between $\langle 1-T\rangle$ and $\langle C\rangle$ and the other three observables. However the $\chi^2$ values for these two fits are very large compared with the number of degrees of freedom, indicating a poor quality of fit. At NNLO the results are more scattered and the values of $\alpha_0$ required are larger. Again, the reduced $\chi^2$ values for the NNLO fits are generally worse than those for the NLO fits. 

For NLO $\overline{\text{MS}}$ PT, in Fig. \ref{scatter single fit MSbar}, there is good agreement in $\alpha_0$ but the error bars are large.  As before the uncertainties decrease at NNLO, as do the reduced $\chi^2$ values. From the $\chi^2$ values in Tables \ref{scatter single fit ECH} and \ref{scatter single fit MSbar} we see that at NLO the quality of the fits using the ECH method is comparable with or better than those using $\overline{\text{MS}}$ PT. At NNLO $\overline{\text{MS}}$ PT describes the data for $\langle 1-T\rangle$ and $\langle C\rangle$ much better than the ECH method.

In Fig. \ref{scatter kappa0} in Section \ref{simple pcs}, when the simple power correction model is added to ECH at NLO, very good agreement is found between the extracted values of $\alpha_s(M_z)$. The fits for $\kappa_0$ are generally close to zero, again showing that NLO ECH does not require large power corrections; this is an observation that was also made in the DELPHI analysis~\cite{DELPHI}. Slight differences in our fit results from the DELPHI collaboration's paper are most probably due to differences in the data analysed and the fitting programs used. Compared with the results for pure ECH from Fig. \ref{scatter alpha} we see that adding a simple power correction brings the values of $\alpha_s(M_z)$ extracted from the outliers $\langle B_W \rangle$ and $\langle Y_3\rangle$ into better agreement, but also increases the unweighted average of $\alpha_s(M_Z)$. At NNLO the grouping is less good and the average of $\alpha_s(M_z)$ is increased even further.

The fitted values of $\kappa_0$ when using $\overline{\text{MS}}$ PT, shown in Fig. \ref{scatter kappa0 MSbar}, are also generally close to zero. The corresponding values of $\alpha_s(M_z)$ have therefore not altered substantially from when pure $\overline{\text{MS}}$ PT (see Fig. \ref{scatter alpha}) is used. The values of $\chi^2$ for the ECH method and $\overline{\text{MS}}$ PT are very similar, implying that the descriptions of each observable given by the two methods are comparable.

\section{Conclusions}\label{Conclusions}

In this paper we have extended the NLO analysis of $e^+e^-$ event shape means performed by the DELPHI collaboration in \cite{DELPHI}, which showed that good agreement with experimental data is achieved over a wide range of energies by using the ECH approach. We had anticipated that applying the ECH method at NLO to higher moments of event shapes would result in similarly good agreement with data, and that extending the analysis to NNLO would improve on the NLO results. Surprisingly, it appears that neither of these expectations is realised in practice. Higher moments are not described well by ECH at NLO, and the NNLO case is even worse. The previously good agreement with data for the means deteriorates when the NNLO corrections are added to the ECH method.

In Section \ref{rho(R)} we attempted to understand the disappointing performance of the ECH method at NNLO by studying the relative magnitudes of the successive terms in the ECH $\beta$-function $\rho({\mathcal{R}})$ in \eqref{dR/dlnx}, as shown in Table \ref{rhoterms}. For the means of event shape moments the NNLO $\rho_2  {\mathcal{R}}^2$ term was the smallest in absolute magnitude, indicating that NLO is the optimum order of truncation. Without calculating the next uncalculated $\rho_3 {\mathcal{R}}^3$ term and showing it to be larger in magnitude than the NNLO term, we cannot definitively say that NLO is the optimum order of truncation. However, the good agreement of the NLO ECH predictions with the experimental data for the means supports this interpretation. For the higher moments the NNLO term is larger in magnitude than the NLO $c {\mathcal{R}}$ term, indicating leading order as the optimum order of truncation. The $Q$ evolution of  ${\mathcal{R}}(Q)$ is controlled by the RS invariant dimensionful constant $\Lambda_{\mathcal{R}}$ of \eqref{Lambda twiddle} which involves the NLO perturbative coefficient $r\equiv r_1^{\overline{\text{MS}}}(\mu_R=Q)$. This implies that leading order truncation is not physically meaningful. 

Another problem in applying the ECH method to the higher moments is the high values of $\Lambda_{\mathcal{R}}$ for these observables. There is a Landau pole at $Q=\Lambda_{\mathcal{R}}$ where the integrated $\beta$-function equation diverges. We have tabulated the $\Lambda_{\mathcal{R}}$ values in Table \ref{lambdaRtable}. For the means the experimental data are at values of $Q$ that are much larger than $\Lambda_{\mathcal{R}}$, which is $\mathcal{O}(1)$ GeV, and so applying the ECH method is effective. For higher moments $\Lambda_{\mathcal{R}}$ becomes of the same order as the energies analyzed, indicating that the method cannot be applied to this data. 

If $Q$ is sufficiently large these problems are avoided as the data will be far away from the pole at $Q=\Lambda_{\mathcal{R}}$. Also, as $Q$ increases ${\mathcal{R}}(Q)$ will decrease due to asymptotic freedom, and so the optimum order of truncation will become NLO or NNLO. It will therefore be possible to apply the ECH analysis to the higher moments at sufficiently high energies. This is of conceptual rather than practical interest, since at experimentally accessible energies the $Q$-evolution of the higher moments is not adequately described by the truncated $\rho({\mathcal{R}})$ function. 

As an extension of this work it would be interesting to approximate missing higher orders using an all orders renormalon resummation matched to the fixed order NNLO result for the $\rho(\mathcal{R})$ function. Such a resummation could be done using a dressed gluon exponentiation~\cite{Gardi} for instance.

Although the ECH method at NLO gives good fits to the means at all values of $Q$ we have investigated possible non-perturbative effects using the dispersive model of Ref~\cite{Luisoni, DokshitzerWebber, SalamWicke}, which has been widely used in other analyses of event shapes, and a simple power correction model used in Ref~\cite{DELPHI}. To include dispersive power corrections in the ECH predictions we have taken the renormalisation scale in the non-perturbative power correction to be that appropriate to the ECH scheme. When performing fits for NLO ECH plus dispersive power corrections we found consistent values of $\alpha_s(M_Z)$ and $\alpha_0$, except for the observables $\langle 1-T\rangle$ and $\langle C\rangle$. This is an improvement on the fit results gained from NLO $\overline{\text{MS}}$ PT plus power corrections. However ECH at NNLO again performs less well than the $\overline{\text{MS}}$ PT counterpart. ECH at NLO gave good agreement in $\alpha_s(M_Z)$ when simple power corrections were added. Generally, only small values of $\kappa_0$ were required, as expected. We conclude that adding power corrections does not improve the performance of the ECH method at NNLO, and that non-perturbative power corrections are not required by ECH to describe the means at NLO. 

Neither the ECH method or standard $\overline{\text{MS}}$ PT provides a good description of the data for
higher event shape moments. It is therefore not clear that one should be recommended for use over the other. However, the excellent results obtained when applying ECH to the means at NLO indicates that further study of this area is desirable.

\subsection{Acknowledgements}

We would like to thank Klaus Hamacher for answering our many questions about the DELPHI data analysis at the start of this project. Peter Richardson and David Grellscheid are thanked for crucial assistance in using \mbox{HERWIG++} to correct the data for b and c decays. K.E.M. gratefully acknowledges the receipt of an STFC studentship.

\bibliography{echpap}{}
\bibliographystyle{h-physrev3}

\end{document}